\documentclass{pasj00}

\begin{document}
\SetRunningHead{S. Deguchi et al.}{Stellar Maser Sources without IRAS Counterpart}
\Received{2004/01/07}
\Accepted{2005/09/10}

\title{ 
Observations of Stellar Maser Sources with no IRAS Counterpart
}

\author{Shuji \textsc{Deguchi}%
  }
\affil{Nobeyama Radio Observatory, National Astronomical Observatory\\
and Department of Astronomical Science, the Graduate University for Advanced Studies \\
Minamimaki, Minamisaku, Nagano 384-1305}

\author{Jun-ichi \textsc{Nakashima}\thanks{Current address: Academia Sinica, Institute of Astronomy
and Astrophysics, PO Box 23-141, Taipei, 106, Taiwan.}
}
\affil{Department of Astronomical Science, the Graduate University for Advanced Studies\\
Minamimaki, Minamisaku, Nagano 384-1305 \\  and \\
Department of Astronomy, University of Illinois 
at Champaign-Urbana, \\ 
1002 West Green St., Urbana, IL 61801, USA}

\author{Takashi \textsc{Miyata}}
\affil{Kiso Observatory, Institute of Astronomy, The University of Tokyo\\
Mitake, Kiso, Nagano 397-0101}

\and
\author{Yoshifusa \textsc{Ita}\thanks{Current address: 
Institute of Space and Astronautical Science, Japan Aerospace Exploration Agency, 
3-1-1 Yoshinodai, Sagamihara, Kanagawa 229-8510.}
}
\affil{Institute of Astronomy, The University of Tokyo\\
Mitaka, Tokyo 181-0015}


%


\maketitle

\begin{abstract}
We investigated stellar maser sources with no IRAS counterpart
at the radio, middle-infrared, and near-infrared wavelengths. 
A 43 GHz SiO maser search for 120 2MASS/MSX objects, and
10 OH 1612 MHz sources with no or a very faint MSX counterpart,
resulted in 43 SiO detections: one OH 1612 MHz source, 
2 near-infrared stars, and 40 MSX sources.
Additional near-infrared $J$-, $H$-, and $K$-band observations of the OH 1612 MHz sources 
detected 5 near-infrared counterparts. Furthermore,
middle-infrared imaging observations at 8.8, 9.7, 12.4, and 24.5 $\mu$m with the Subaru 8.2-m telescope  
found counterparts for 2 near-infrared stars with SiO masers,
and counterparts for 6 OH 1612 MHz sources.
However,  4 OH 1612 MHz sources were not detected
in the sensitive near- and middle-infrared searches; 
three of these are relatively strong OH
maser sources for which the positions were known accurately.
We conclude that one of these (OH 028.286$-$01.801) must be a young object 
in a star-forming region.
\end{abstract}

\section{Introduction}

It has been known that the IRAS Point Source Catalog (PSC) is incomplete 
in the Galactic-center area, and in the strip (the IRAS gap),  
which traces a great circle almost perpendicular to the galactic plane 
at $l=84^{\circ}$ (\cite{bei88}). 
The cataloged source densities in these areas are apparently 
lower than those in the surrounding regions. Because past maser surveys, especially SiO maser surveys,
were often made based on the IRAS PSC (for example, \cite{deg00a,nak03a}),  
some important objects may have been left unobserved.
A lack of maser surveys in these areas caused an incompleteness in the
SiO maser statistics of galactic-plane sources (for example, \cite{jia99}). 
In the area near the Galactic center,
an effort has been made to fill the unsurveyed area (\cite{deg04a}) by selecting
large-amplitude variables identified in infrared bands (\cite{gla01}).
However, the situation has now been changed owing to
the MSX survey (\cite{ega99}), which cataloged middle-infrared (MIR) sources 
in the IRAS-incomplete regions.  It covered the whole galactic plane with 
$|b|\lesssim 5^{\circ}$, and partially the IRAS-gap area. Furthermore, the 2MASS
catalog (\cite{skr00}) became available for selecting mass-losing Asymptotic Giant Branch (AGB ) 
candidate stars, which are suitable for maser searches.

In this paper, we present the result of an SiO maser search of 
IRAS-missing bright objects in the galactic plane, and objects in the IRAS-gap area
near $l= 84^{\circ}$. Furthermore, we studied the nature 
of OH 1612 MHz sources in the galactic plane 
in the near- and middle-infrared bands. 
This work was originally motivated by an accidental detection of 
SiO masers at the 10$'$-offset position of 
IRAS 18268$-$1117 (=OH 020.433$-$00.344), where
no corresponding infrared object was found
at the time of the detection in 2001 March
(a short summary of this detection was given in Appendix 1).
Therefore, we planned to search for SiO maser sources
without near- or middle-IR counterparts.
However, gradual releases of the 2MASS catalog and the
MSX5C/6C catalogs enabled us to identify the above source as a faint MIR object.
We made sensitive near- and middle-IR imaging observations for a few selected maser
sources to check whether or not these sources have counterparts in the relevant infrared bands.
We found  near- and middle-IR counterparts for all of the 
SiO-detected sources. We found that there are several OH 1612 MHz sources 
having no near- and middle-IR counterpart. We describe the results of these observations and 
discuss the nature of the odd maser sources without
near- or middle-IR detections.

\section{Radio Line Observations}

Observations in the SiO $J=1$--0 $v=1$ and 2 maser lines around 43 GHz were made 
with the Nobeyama 45-m telescope
during the periods in 2001 March--April and 2002 February--May. Some additional observations 
were made in 2004 May and 2005 January--February, after the MSX6CG catalog out of the galactic plane was available.
A cooled SIS-mixer receiver (S40) with a bandwidth of about 0.4 GHz were used, 
and the system temperature (including atmospheric noise) was 200--300 K (SSB), 
depending on the frequency and weather. 
The aperture efficiency of the telescope at 43 GHz
was 0.5. The antenna temperature
given in the present paper was corrected for atmospheric absorption
and telescope ohmic loss, but not for the beam or aperture efficiency  ($\equiv T_{a}^*$).
The conversion factor of the antenna temperature to the flux density is about 2.9 Jy K$^{-1}$.
Acousto-optical spectrometer arrays of both high and low resolutions 
(AOS-H and AOS-W) with 40 and 250 MHz bandwidths, respectively, were used,
giving velocity coverages of about $\pm 350$ km s$^{-1}$
and spectral resolutions of 0.3 km s$^{-1}$ (per two binned channels).  
The telescope pointing was checked using strong SiO 
maser sources, IRC$-$10414 and V1111 Oph, and the pointing accuracy was confirmed to be
better than 5$''$ in a windless condition. 
Observations were made in a position-switching mode, 
and the off-position was chosen to be 10$'$ away from the object in right ascension.
The total on-source integration time was typically 10 min. 
We also observed several sources in the H$_2$O maser line at 22.235 GHz 
using a HEMT (H22) receiver. This was made mainly as a backup observation
in bad weather. The aperture efficiency at this frequency was 0.62;
the conversion factor to the antenna temperature was 3 Jy K$^{-1}$.
The HPBW was 73$''$. The system temperature at 22 GHz varied between 180--400 K,
depending on the weather.

The line parameters of SiO detections are given in table 1
and the rms noise levels of non-detections are given in table 2.
The infrared properties of the observed sources are summarized in table 3,
listing the 2MASS name, $K$, $J-H$, and $H-K$ magnitudes, MSX 6C name, separation from the
2MASS positions, flux density in C band (12.3 $\mu$m), colors $C_{\rm AC}$ [$\equiv log(F_{\rm C}/F_{\rm A})$] 
and $C_{\rm CE}$ [$\equiv log(F_{\rm E}/F_{\rm C})$], and nearby IRAS source name within $1'$, and separation
in arcsec. 

We selected two types of objects: 
one from 2MASS NIR objects with no IRAS (or very faint MSX) counterpart, 
and another from OH 1612 MHz sources without IRAS 
or MSX counterpart. In addition, we observed bright MSX sources without IRAS identifications 
near the Galctic plane, especially in the IRAS gap-area.

\subsection{SiO Observations for 2MASS/MSX and OH 1612 MHz objects}

We selected the observing candidates using
the 2MASS Gator by the $J-K$ color and the $K$ magnitude ($J-K>3.0$ and $K<8.0$).
It was checked whether or not the selected candidates had 
IRAS or MSX(5C) counterparts; the stars that had an IRAS or MSX counterpart 
within 20$''$ (a radius of HPBW of the 45-m telescope) were removed 
from the list. We observed 32 objects from this sample.
The positions of the observed sources and their near-infrared (NIR) properties
are listed in table 3 with ``$\sharp$'' after the 2MASS name in the first column. 
Because of a software bug of the MSX(5C) imager, 
which was used for source selection, the objects listed
in table 3 are not perfectly consistent with the above condition.
However, all of the objects have very low MIR flux densities.  
   
Another type of object without MIR counterparts 
is a subsample of OH 1612 MHz sources in the galactic disk
($l<45^{\circ}$; \cite{sev01}). These sources
were detected with VLA, and the positions are 
known within a few arcsec accuracy.
The assignments to the nearest IRAS/MSX/2MASS
sources were made by \citet{sev01}.
A considerable number of the objects
have IRAS/MSX counterparts at positions beyond 3 sigma of the
IRAS/MSX-position uncertainties. 
Because the assignments to the MSX sources
by Sevenster et al.'s (2001) work were made before the software-bug correction 
of the MSX imager, we again checked whether
they really have no MSX counterpart. Finally,
we chose 10 OH 1612 MHz objects with no IRAS nor
MSX counterpart (except OH 006.095$-$00.630, 013.379+00.050, and  014.431$-$00.033, 
which were included in the sample before the software bug correction). The positions and properties 
of these OH objects are summarized in table 4, listing the OH name,
positions, the center velocity of OH double peaks, the outflow velocity,
the MSX6C source name, the flux density at the A band, the separation from the OH position,
the nearest IRAS source name (and separation in arcminute between the parentheses),
and the object type. 

We detected SiO maser emission in OH 006.095$-$00.630, 
and H$_2$O in OH 28.397+0.080.
The SiO and H$_2$O spectra for these two objects are shown at the bottom 
of Figure 1b.
Weak water maser emission in OH 28.397+0.080 can be seen
on the lower velocity side ($V_{\rm lsr}=15$--60 km s$^{-1}$) 
of the main peak at $V_{\rm lsr}=77.2$ km s$^{-1}$. The nature of these objects
is discussed in sections 4.2 and 4.3.


\subsection{SiO Observations in the IRAS Gap.}

The IRAS gap area of ($l$, $b$)$\sim$(80$^{\circ}$--88$^{\circ}$, $-6^{\circ}$--6$^{\circ}$) is 
an easily accessible region of the sky from Nobeyama. 
However, past SiO maser surveys 
(\cite{jia97,nak03a}) did not involve objects in
this region.
The MSX survey was mainly restricted to the
galactic plane ($|b|\lesssim 5^{\circ}$), except for
a few areas out of the galactic plane, which was
cataloged as ``High latitude, $|b|\gtrsim 6^{\circ}$'' in MSX 6C. 
From the MSX catalog, we selected MIR
sources above the 5 Jy at band C (12.13 $\mu$m). 
Because the MSX survey did not cover 
the 60  $\mu$m band, source selections were made 
simply based on a condition for the logarithmic flux-density ratio, 
$-0.6<C_{\rm CE} \equiv$ log($F_{E}/F_{C})<0.6$,
where $F_{C}$ and $F_{E}$ are the flux densities at band C (12.13 $\mu$m) and band E (21.3 $\mu$m).
The logarithmic flux-density ratio,  $C_{\rm CE}$,
is very similar to the IRAS color, $C_{12}$, which is the logarithmic 
ratio of the IRAS 25 to 12 $\mu$m flux-density. 
The positions and MIR properties of the observed sources
are listed in table 3 (without ``$\sharp$'' in the first column). 
Because most of these objects have 2MASS NIR counterparts,
we designated the objects by the 2MASS counterpart names.
We also observed MSX objects with IRAS counterparts and a few bright Miras 
in the same sky area for completeness; these objects
have colors of AGB stars and were not observed before.

We detected SiO masers 39 out of 86 objects in  MSX sources near the galactic plane, or
26 out of 62 in the IRAS gap region of $70^{\circ}<l<92^{\circ}$.
The results are also given in tables 1 and 2. Because these objects are mostly
bright MIR objects, discussions on the overall properties of these sources are given 
in appendix 2. One of the most interesting objects among SiO detections is
a symbiotic star, J21020980+4546329 (=V407 Cyg; \cite{tat03}). 
This star has been known to be similar to OH/IR sources (\cite{mun90}). However,
it seems to have escaped from past sensitive maser-line searches
because of not being included in the IRAS PSC, except for a few negative results (\cite{pat92,sea95}).
The detection of SiO masers in the present work indicates that this is
a star similar to another symbiotic star with SiO masers, R Aqr.   

\section{Infrared Observations}

\subsection{Near-Infrared Observations}

Near-infrared imaging observations for the OH 1612 MHz objects 
were made during 2001 August 25--27  
with the 88-inch telescope of University of Hawaii, at Mauna Kea, Hawaii, 
using the NIR array camera, called SIRIUS 
(Simultaneous-color InfraRed Imager for Unbiased Survey). 
The infrared array camera had 1024$\times$1024 pixels (HgCdTe)
with a resolution of $0''.28$ per pixel, and obtained  
 the $J$- (1.25 $\mu$m), $H$- (1.63 $\mu$m), 
and $K$-band (2.14 $\mu$m; in fact, $\equiv K_{s}$) images simultaneously.
This camera was developed by the SIRIUS team at Nagoya University 
for a Southern Sky Survey with the 1.4-m infrared telescope 
at Southerland, South Africa.
It was temporarily installed to the UH 88-inch telescope
for a joint UH--NAO (National Astronomical Observatory) project.
A detailed description of this camera can be found
in \citet{nag99} or \citet{nag03}. 

$JHK$-band images for programmed objects were 
taken with an exposure time of 30 s at two dithering positions 
in order to subtract the sky background. 
The data reductions were made by a standard data-reduction procedure with the IRAF package, 
subtracting the dark frames, dividing by the sky background, and
 subtracting the sky-field. Then, $JHK$ color-composite images were created.
Objects with ``red'' colors were searched in the color-composite images.
These objects normally fall within the regions $J-H>1$ and $H-K>0.5$ on 
a NIR two-color diagram; we have  long experience of identifying OH/SiO objects 
(see \cite{deg98,deg01a}).
The photometric results for OH sources are summarized in table 5, listing 
the source name; the positions of the identified star
in J2000; the $J$, $H$, and $K$ magnitudes; and whether or not SiO/H$_2$O is detected,
and whether or not the source is recognizable on the MSX imager.
We identified 5 NIR counterparts near the OH positions (see figure 2).
However, for the other 4 objects, we could not find any counterparts.
For the case of OH 006.095$-$00.650, a very faint red star was detected 
(only in $K$ band) near the OH position (within 2$''$). However, because of
the faintness (and probably due to time variation and sensitivity),
this star is not seen on the corresponding 2MASS $K$-band image.

\subsection{Middle-Infrared Observations}

Middle-infrared observations were made on 2002 July 29
using the Subaru 8.2-m telescope of National Astronomical Observatory of Japan
at Mauna Kea, Hawaii. The MIR camera, 
COMICS (COoled Mid-Infrared Camera and Spectrometer), was
mounted on the Cassegrain offset focus. This camera has a $320\times 240$
Si:As detector array, and it was optimized for $N$- and $Q$-band
observations.  The camera
provided a field of view of $42'' \times 31''$ with the pixel scale of $0''.13$.
The details of this camera are
described in \citet{kat00}, or the Subaru instrumental home page.\footnote{
$\langle$http://SubaruTelescope.org/Observing/Instruments/COMICS$\rangle$}

Observations were made in the chopping mode by a second mirror
for background subtraction at a 30$''$ offset position.
In addition, dithering of about 5$''$ was also introduced.  
These techniques to subtract the sky background 
reduced the noise by about $\surd 2$ compared with simple staring.
We used a camera with 3 filters [ 
N8.8 (8.78 $\mu$m with a band width of 0.77 $\mu$m),
N9.8 (9.73$\mu$m with a band width of 0.94 $\mu$m),
and N12.5 (12.41$\mu$m with a band width of 1.15 $\mu$m)],
with exposure times of a few to 30 s.
The flux standard stars (HD 161096, HD 168723, HD 169916 etc.) 
were observed as a flux calibration. The seeing size of a point source
was typically $0''.35$ at 12.4 $\mu$m.
The raw data were reduced using a software package developed by the COMICS team, 
with subtracting the dithering images and dividing by sky-flats.
We used the IRAF package for aperture photometry.
Because we normally detected only one source in the field 
and it was near the center of the field,
the photometry was extremely simple. 
We used an aperture having a radius of 6 pixels,
and integrated all of the emission inside. To determine the
sky background emission, we took the average 
count between two radii of 20 and 30 pixels from the star 
and subtracted the average from the count in the aperture. 
For nondetections, we measured the level of the blank sky at several different positions
in the same manner as above and gave the upper limit of the signal
as 3-times the rms value of the counts. 
Two exposures in each band were normally made for one object.
The average flux densities calculated from these exposures are given in table 6.
The errors in the flux densities were estimated from the difference 
of the flux densities in two exposures, and were expected to be about 0.1 Jy.
The absolute flux densities were obtained by comparing the flux densities with those 
of the standard stars at nearly the same elevation (\cite{coh99}).

The Q-band imaging observations at 24.5 $\mu$m became available
with Subaru COMICS after the 2003 first semester. Therefore, additional imaging observations 
were made at 24.5 $\mu$m on 2003 July 15 as a service observation. At this time, we chose 5 objects 
that were not detected (or marginally deteted) before in the other bands.
The data reduction was made in the same way as described before.
The rms noise level was about 0.5 Jy. We detected only OH 18.381+00.162
at 24.5 $\mu$m.


The results are summarized in table 6; the source name, observed flux densities at
8.7, 9.7, 12.4, and 24.5 $\mu$m, the dust color temperature ($T_{\rm dust}$), the color ($C_{12}$),
$K$-band extinction ($A_{K}$), estimated luminosity distance ($D_{L}$), and
kinematic distance ($D_{\rm K}$) are given.
The dust color temperature was computed from a flux density ratio
of 12.4 to 8.7 $\mu$m assuming a black body; 
the color, $C_{12}$ [$\equiv$ log($F_{25}/F_{12}$), where $F_{12}$ and $F_{25}$
are the flux densities at 12 and 25 $\mu$m],
was computed from the dust color temperature (of 8.7 and 12.4 $\mu$m) 
assuming that the dust emits black body radiation
at 25 $\mu$m. The kinematic distance was estimated 
from the OH radial velocity, assuming a flat rotation curve
of 220 km s$^{-1}$ and a Sun--G.C. distance of 8 kpc. 
As far as the color temperatures (or $C_{12}$) are concerned, 
the stars that were detected at 8.7  and 12.4 $\mu$m
were not different from typical OH/SiO maser sources (e.g., \cite{nak03a}).
The $K$-band interstellar extinction and the distance were estimated using the same formula
as that of \citet{deg02}, assuming the extinction law in the exponential-disk model in the Galaxy
and a constant stellar luminosity of $8 \times 10^3 \; L_{\odot}$ of the central star.


In the case of J18301610$-$1115376, the MSX catalog gives flux densities
of 1.72 and 2.10 Jy at  8 and 12 $\mu$m, respectively 
(for MSX6C\_G020.5138$-$00.4885). In the case of OH 006.095$-$00.630, the MSX catalog 
gives flux densities of 1.59 and 2.48 Jy at  8 and 12 $\mu$m, respectively
(for MSX6C\_G006.0952$-$00.6294).
The observed values were roughly 20--30\% larger than the flux densities
given in the MSX catalog for these sources. However, because
the MIR fluxes of pulsating stars are variable (for example, \cite{miy00}),
the differences can be attributed to the time variation. We conclude that
they coincide reasonably well with the previously measured flux densities for these two stars.


\section{Discussion}
In this section, we discuss the properties of the NIR and 
OH 1612 MHz sources.
Because the properties of MSX sources in the area of $l\sim 84^{\circ}$
are quite similar to the normal IRAS AGB objects, we give a discussion
on them separately in appendix 2. 

\subsection{Characteristics of NIR and MIR Spectra}
Figure 3 shows two-color and magnitude--color diagrams for the observed 32
2MASS objects with no IRAS or a faint MSX counterpart ($20^{\circ} <l<80 ^{\circ}$). 
We can recognize that the distributions of the sampled stars 
in these diagrams are quite similar to those of the corresponding diagrams of SiO detected objects
in the galactic disk and bulge (\cite{deg01a}, 2002, and 2004b),
except that the $K$ magnitudes of the present sample
are about 0.5 magnitude on average higher (fainter) than those of the outer bulge objects. 
This fact indicates that the sampled NIR objects
are approximately 8 kpc or more away from the Sun. The two objects detected
in SiO are relatively bright compared with the other nondetection stars,
indicating that they are relatively close to the Sun ($D_{L} \lesssim 8$ kpc). The 
MIR flux densities of these objects seem to indicate that
they have a smaller mass-loss rate than those of the average SiO maser stars
in the galactic disk/bulge do. Therefore, the SiO detection rate
of these stars is considerably lower than that of the IRAS sources in the disk, 
partly due to their distances.
The other reason for the low detection rate is attributed to the fact 
that this sample involves the RGB or AGB stars without mass loss,
because the sampling of NIR stars was governed only by the NIR colors.

Figure 4 shows the observed spectra of the detected
sources between 1.25 and 12.4 $\mu$m. The upper panel shows the
spectra of near-IR sources with SiO, and the lower panel shows the spectra
of OH 1612 MHz sources. It has been well known that the silicate dust has
an emission/absorption band at 9.8 $\mu$m. 
The silicate band exhibits emission at 9.8 $\mu$m
when the circumstellar envelope is optically thin, and an absorption
when it is optically thick [for example, see \citet{miy00}]. 
The absorption feature is recognizable for some sources in figure 4
(clearly in OH 006.095$-$00.630).
However, the emission feature is not very clearly
recognized in figure 4 (except a slight enhancement in OH 031.985$-$00.177). 
This is probably because the emission feature
at 9.8 $\mu$m is smeared because of the wide band width ($\sim$0.94 $\mu$m) 
of the N9.8 filter. In contrast, the absorption feature can be  
wide, and most of the OH sources exhibit the 9.8 $\mu$m absorption 
except OH 031.985$-$00.177. 

In figure 4, the near-IR sources with SiO masers exhibit flat spectra between 2.2 and 12.4 $\mu$m.
The circumstellar envelopes of these sources  are relatively thin.
However, OH 013.510$-$00.578 exhibits a relatively flat spectrum, except for 
a weak absorption feature at 9.8 $\mu$m. This source is probably
a star similar to the near-IR objects with relatively thin envelopes,
but it must be located far (by a factor of 3 compared with near-IR sources).

\subsection{OH 018.381+00.162 and OH 032.731$-$00.327}
It has been known that the OH 1612 MHz masers are
occasionally found in star-forming regions, accompanying the dominant 
OH 1665/1667 MHz masers [for example, \citet{gau87}]. The positions
of these main- and satellite-line masers usually coincide within a few arcsec accuracy.
In fact, \citet{cas99} found that 
at least 7 of the 1612 MHz sources (mostly single-peak spectra except one)
that were found by \citet{sev97b}
are associated with star-forming regions. Among them, the object,
OH 5.885$-$0.392, has a doubly peaked spectrum with a separation of 40 km s$^{-1}$,
exhibiting a stronger 1612 MHz (peak) flux density than the 1665 MHz flux density.
Because of the association with a continuum source,
this object is considered to be a star-forming region (\cite{zil90}).
On the other hand, \citet{cas99} discussed 5 OH 1612 MHz sources [in table 2 of \citet{cas99}],
which are seen toward, but are not associated with, star-forming regions;
all of these have IRAS or MSX counterparts with reasonable $C_{12}$ colors as evolved stars. 
The 2MASS images show only one clear NIR counterpart   
(OH 331.646$-$0.259), but no or a dubious red candidate 
for the other 4 objects. One object, OH 331.594$-$0.135,
 has no IRAS, MSX, or 2MASS counterpart.

These examples tell us that, even if we cannot find any NIR counterpart
for a particular object, we should often find the MIR counterpart
if it is an evolved star (in AGB or post-AGB phase). If not, they may be a star-forming region.
A MIR and NIR diagnostics, as well as that using the OH main line (1665/1667 MHz) 
and SiO masers, is crucial for judging whether the object is an evolved star 
or a star-forming region, and is useful for precluding 
non-AGB stars from OH radial-velocity samples in the galactic disk (\cite{deg04b}).

In the present sample of OH 1612 MHz objects, 
SiO masers are detected only in OH 006.095$-$00.630.
The presence of the NIR and MIR counterparts secures that this object
is an evolved star.
As far as we find the bright
NIR counter parts, as for OH 013.379+00.050, 013.510$-$00.578, 014.431$-$00.033, and 031.985$-$00.177,
they are likely to be evolved stars.
Furthermore, the NIR counterparts of these OH objects do not accompany any 
star-forming activities, such as nebulous features 
and a cluster of faint red stars, which are often associated with star-forming regions.
On the other hand, for the other 5 OH 1612 MHz objects for which neighther NIR nor MIR
counterparts are found (except OH 18.381+00.162),
they are possibly OH maser sources in star-forming regions.

Two sources, OH 018.381+00.162 and 032.731$-$00.327,
are relatively strong 1612 MHz emission sources
among the sampled OH sources: about 4 Jy km s$^{-1}$ 
for the blue shifted components of the double peaks. The 
expansion velocities are about 9 and 15 km s$^{-1}$, respectively.
Since the OH peak flux density is roughly correlated with the
IRAS 25 $\mu$m flux density (\cite{tel91}), we estimate that the 25 $\mu$m
flux densities of these sources must be about 15--60 Jy if they are AGB stars. 
The weakness of the 12 $\mu$m emission of these objects indicates $C_{12}>1$.  
The positions of OH 018.381+00.162 
were measured using VLA (\cite{bow83}; \cite{fix84}),
and both positions coincide well with the later measurement (\cite{sev01}) 
with an error of $\sim 1''$. 
Fix and Mutel (1984) found that the OH emission of OH 018.381+00.162 was not resolved
with VLA and gave the lower limit of the distance, 7 kpc.
Water maser searches for this star have been negative (\cite{eng02};
the present study). OH 032.731$-$00.327 was not detected before or after
\citet{sev01}. Though a strong OH/IR source, OH 32.8$-$0.3 (=V1365 Aql), 
has OH double peaks at almost the same radial velocities (\cite{win75})
as those of OH 032.731$-$00.327,
with a positional separation of $5'.8$ (\cite{joh77, her85}),
it is not listed by \citet{sev01}.

For comparison, we took an OH/IR source, OH 037.1$-$0.8, 
which is one of the well-known OH/IR objects
(=IRAS 18596+0315) and an H$_2$O/SiO maser emitter (\cite{tak01,eng02,jew91}). 
The OH flux density and expansion velocity are 4 Jy and 14 km s$^{-1}$, respectively.
This object has IRAS flux densities, 2.6, 14.2, and 22.6 Jy
at 12, 25, and 60 $\mu$m, respectively. A steep increase 
of the intensity beyond 12 $\mu$m ($C_{12}=0.73$) indicates that this object
is at the phase of proto-planetary nebulae. The distance of this source
was estimated to be 5.5 kpc from NIR photometry 
(\cite{sun98}). 
If OH sources without NIR/MIR counterparts, 
OH 018.381+00.162 and OH 032.731$-$00.327,
are located as distant as 11 kpc, 2-times the OH 037.1$-$0.8 distance, 
the 12 $\mu$m flux density would be about 0.6 Jy, which is close to
the detection limit of Subaru COMICS. 
In fact, OH 018.381+00.162, was detected at 0.61 Jy at 12.4 $\mu$m,
and 3.5 Jy at 24.5  $\mu$m
in the present work (see table 6). The MSX imager gave a slight
enhancement of signal on the D- and E-band (14 and 21 $\mu$m) images 
at this position. These facts are consistent with the estimated distance 
of about 11 kpc for OH 018.381+00.162. 
For OH 032.731$-$00.327,
we found no enhancement of emission on the MSX A--E-band images 
at the OH position. Therefore, it is more difficult to
estimate the distance.

Because the envelope expansion velocities of these two sources
($\sim 9$ and 14 km s$^{-1}$) 
are as moderate as that of an AGB star, the mass losses 
of these two sources are relatively mild.
\citet{sev02} argued that cold OH/IR sources with $C_{12}>0.5$ have
a luminosity of $10^4 \; L_{\odot}$ and a mass below 4 $M_{\odot}$,
and that they are less luminous and less massive than those
in the bluer region, $0 \lesssim C_{12} \lesssim 0.5$ 
and $0 \lesssim C_{32} \lesssim 0.9$,
in the two-color diagram. The distribution of these cold OH/IR objects
at the left side of the black body curve in the two-color diagram
seems to be fitted by a model of 3 $M_{\odot}$ without dust formation
in the post-AGB phase (\cite{van97}).
The models without dust-formation in the post-AGB phase
seem to be consistent with the nondetection
of SiO masers in these sources.

The detectability of the central star in the $K$ band 
may strongly depend on the sphericity of the envelope in the post-AGB phase.
If a collimated high-velocity flow is developed and a hollow is created, 
the NIR photons can escape through the hollow.
In this sense, $K$-band detection
would be a measure of sphericity of the envelope.
The color, $K-[12]$, in the models of \citet{van97} for cold objects 
falls between 4--10, where [12] is the 12 $\mu$m magnitude 
[$\equiv$2.5 \ log($59.3/F_{12}$)].
The 12 $\mu$m magnitude is estimated to be nearly 5.2 (or larger) for 
these two OH sources. Because these sources  were not detected in the $K$-band
observations, we estimate $K-[12]$ to be about 10 for these sources. 
Our findings given in this paper is consistent with 
Sevenster's conjecture (\cite{sev02}) that the cold objects
are more or less evolved to elliptical planetary nebulae, rather than 
bipolar planetaries. 


\subsection{The Nature of OH 028.397+00.080}
We found that the OH source, OH 028.397+00.080, has
no NIR and MIR counterpart, though
a relatively strong H$_{2}$O emission was found in the present work.
Note that the 1612 MHz spectrum shows only a single peak
at $V_{\rm lsr}=74.9$ km s$^{-1}$ (\cite{sev01}).

Here, we speculate on a few nearby objects that have been found around this OH source.
An MSX point source, MSX6C\_ G028.3937+00.0757 ($18^{\rm h}42^{\rm m}52^{\rm s}.61$, $-04^{\circ}00'12''.5$, J2000.0),
is located about 20$''$ southeast of OH 028.397+00.080.
This MSX source is deduced to be the same as IRAS 18402$-$0403
($18^{\rm h}42^{\rm m}54^{\rm s}.91$, $-04^{\circ}00'13''.1$, J2000.0), where
the IRAS flux densities at 12 and 25 $\mu$m, 1.2 and 16.4 Jy,
correlates well with those of MSX, respectively.
The flux densities at 60 and 100 $\mu$m are 
356.3 and  932.3 Jy, respectively, which are
too large as an evolved object.
The spectral energy distribution of this source
indicates that this is a very cool object, such as 
a young star embedded in a molecular cloud, or a galaxy. In fact, \citet{bec94}
found a radio continuum source at almost the same position
($18^{\rm h}42^{\rm m}52^{\rm s}.81$, $-04^{\circ} 00' 11''.2$, J2000.0) as that of the MSX source;
according to them, this is a
GPS (Gigahertz Peaked Spectrum) radio source. 

During mapping of the molecular cloud, G28.34+0.06
(which was found as a silhouette on MSX middle-IR images), 
\citet{car00} detected a strong submillimeter-wave point source
(assigned as P2 in their figure 3) with JCMT SCUBA at the position 
($18^{\rm h}42^{\rm m}52^{s}.4$, $-03^{\circ}59'54''$, J2000.0). It is
located about 6$''$ west of OH 028.397+00.080, 
indicating a reasonable agreement with the position 
of OH 028.397+00.080 within the errors.
The 850 $\mu$m map shows considerable complexity of this source.
Based on a position difference of about 23$''$,
\citet{car00}  noted that this source is distinctively different 
from IRAS 18402$-$0403 (or MSX6C\_ G028.3937+00.0757  
).
From the high flux density (48 Jy) at 450 $\mu$m, 
they concluded that it is a pre-main-sequence OB star
at a distance of $\sim 5$ kpc. 

If this submillimeter-wave source is an AGB star embedded 
in a very thick envelope, the
luminosity must be about $10^4 L_{\odot}$. 
It is also possible that a considerable part of middle- and far-infrared 
flux densities of the above-mentioned nearby source, 
IRAS 18402$-$0403 ($F_{60}=459$ Jy), may come from this OH 1612 MHz source.
Assuming that half of the 60 $\mu$m flux of IRAS 18402$-$0403
comes from this OH source (a half from dust clouds toward the same direction), 
we obtain a distance of about 6 kpc
from the Sun. The high flux density at 450 $\mu$m
indicates that the dust is considerably opaque
at this frequency. The optical depth of unity
at 450 $\mu$m indicates dust extinction, $A_{K}=7.2$ mag
(equivalent to the column density $\sim 1.3\times 10^{23}$ HI cm$^{-2}$).
This value is much larger than the column density 
$5 \times 10^{20}$ HI cm$^{-2}$  for (slightly unusual)
massive HI cloud of $10^5 \; M_{\odot}$ (\cite{min01}) in this direction.
Therefore, we can consider that most of the submillimeter radiation
(and far-IR emission too) comes from the envelope 
of this star. 
Based on the column density and the
radius of the black body sphere, $r=2.3 \times 10^{16}$ cm and $T=70$ K
(which gives 48 Jy at 450 $\mu$m), the mass loss rate of the star 
can be estimated as $3\times 10^{-4} \; M_{\odot}$ 
when we assume an expanding velocity of 30 km s$^{-1}$. 

This model is considerably extreme as the circumstellar envelope of an O-rich evolved object. 
Therefore, we conclude here that OH 028.397+00.080 is a cool object 
in a star-forming region, not an evolved star.


\section{Conclusion}
We investigated the nature of stars with optically very thin 
or extremely optically thick dust envelopes 
in the radio, NIR, and MIR wavelengths.
By SiO maser lines, we detected 44 out of 130 observed objects.
Infrared observations for the subset of this SiO sample 
confirmed that about 2/3 of the objects with thick envelopes (OH 1612 MHz sources)
have NIR or MIR counterparts. So far, we have found NIR counter parts 
for all of the SiO detected sources. However, we could not find
any NIR or MIR counterparts for 4 of OH 1612 MHz sources. 
These objects must be stars with extremely cold thick envelopes
without dust formation in the AGB phase, 
or young objects in star-forming regions.  

\

The authors thank Dr. A. Winnberg for reading the manuscript and useful comments. 
They also aknowledge Dr. T. Fujiyoshi for help with the Subaru COMICS observations, 
Dr. K. Sugitani and the SIRIUS team
for help with the UH88$'$ SIRIUS observations, at Mauna Kea, Hawaii,
Dr. P. Wood for taking images of J18301610$-$1115376
with the ANU 2.3-m telescope, and Ms. L. Cao of Beijing observatory 
for data reductions. This research was partly supported by a Grant-in-Aid for
Scientific Research from Japan Society for the Promotion of Science (C1260243).

\section*{Appendix 1. SiO Detection toward J18301610$-$1115376.}
This work was originally motivated by an accidental detection of 
SiO masers at a 10$'$-offset position from 
IRAS 18268$-$1117 (=OH 020.433$-$00.344) on 2001 March 23, where
no corresponding infrared object was found.
The subsequent 5-point mapping of this SiO emission 
(HPBW of 40$''$) on 2001 April 10 gave the accurate position 
of this source as (RA, Dec, epoch)= 
($18^{\rm h}30^{\rm m}15^{\rm s}.9$, $-11^{\circ}15'35''$, J2000.0);
the positional accuracy was estimated to be about 5$''$.
We found no IRAS point source within 2$'$. We also checked
all of the known optical, infrared and OH 1612 MHz objects 
in the SIMBAD database, and found no candidate for this source.

A subsequent near-IR imaging observation at this position 
was kindly made by P. Wood with the ANU 2.3-m telescope on 2001 June.
The $J$- and $K$-band images showed a bright-red star at 
 (RA, Dec, epoch)= ($18^{\rm h}30^{\rm m}16^{\rm s}.10$, $-11^{\circ}15'33''.0$, J2000.0).
Because of the brightness at the $K$-band,
$\sim 5.8$ mag and the $J-K$ color, $\sim 3.6$ mag, 
it must be, without doubt, a near-IR counterpart 
of the above SiO maser source. We designated this object
as J18301610$-$1115376 for the 2MASS corresponding object,
after 2MASS data was released.

The MSX catalog listed a source, 
MSX5C\_G020.5046$-$00.4779 (0.5 Jy in band A [8.8 $\mu$m], 
but not detected in the other bands),
which is located about 45$''$ south of the SiO position.
This source 
is too weak and cannot be a strong SiO maser source
as above. For confirmation, we observed this source 
(=J18301278$-$1115463=MSX6C\_G020.5053$-$00.4777 in table 2 and 3)
in the SiO $J=1$--0 $v=1$ and 2 transitions on 2001 April 10;
no emission stronger than 0.08 K was detected.
Meanwhile, it turned out that the original MSX catalog
and Image Server (ver. 1) had a software bug, 
and did not list some faint objects below a few Jy. 
The MSX Image Server and Overlays (ver. 2)\footnote{
$\langle$http://irsa.ipac.caltech.edu/applications/MSX/ $\rangle$.} 
gave MSX6C\_G020.5138$-$00.4885 at
 (RA, Dec, epoch)=(18$^{\rm h}30^{\rm m}16^{\rm s}.10$, $-11^{\circ}15'37''.1$, J2000.0)
with 2.2 Jy in band C (12.5 $\mu$m). This MSX source is
located about 4$''$ SE of above SiO position, and 
therefore, we believe this is a MIR counterpart 
of the above-mentioned SiO maser source.

\section*{Appendix 2. MSX Sources in the Region of $70^{\circ}<l<92^{\circ}$.}

We detected 26 among 62 2MASS/MSX sources observed in SiO maser lines
in the region of $70^{\circ}<l<92^{\circ}$,
where the IRAS survey was incomplete.
The SiO detection rate is about 42 percent, which is slightly smaller than the
SiO detection rates in the previous bulge/disk SiO surveys
(\cite{izu99,deg00a,nak03a}). 
Figure 5 shows histograms of 12 $\mu$m (band C) flux densities, $F_{12}$ 
and the colors, $C_{\rm CE}$, for the observed objects.
The solid line in these figures indicates the SiO detection rate in each bin.
The detection rate looks nearly constant over log($F_{\rm C}$), except at low and high flux-density limits
where the statistical significance is poor. The detection rate tends to increase
with decreasing $C_{\rm CE}$. However, this is probably due to a deficiency of
high $C_{\rm CE}$ objects in this galactic-longitude area compared with objects in
the inner-Galaxy areas (for instance, see \cite{jia99}).

It has been known that the SiO detection rate of IRAS sources drops at
large $C_{\rm CE}$ (for example, see \cite{nak03b}). 
Though classification of the MSX sources have been made
(\cite{lum02}), comparisons between the MSX and IRAS sources,
especially for sources with the colors of AGB stars, have not been investigated well.
\citet{jia02} discussed that the SiO maser detection rate is correlated
with the MSX band A (8 $\mu$m), because the SiO molecule has the 
fundamental vibration band at around 8 $\mu$m.
The detection rate in the present sample was comparable with
those expected at this longitude range. \citet{jia96} discussed that
the SiO detection rate for the IRAS sample decreases with 
the galactocentric distance. For example, it was 13\% for the sources
in $l=90$--250$^{\circ}$, and 42\% for the sources in 
$l=55$--90$^{\circ}$. They concluded 
that the contamination rate by C-rich or young stars 
in the flux-limited sample increases with the galactic longitude
(\cite{jia99}). The present SiO maser search gave 
a detection rate similar to the previous one made in the region of $l=55$--90$^{\circ}$.
A small improvement for the source selection was made in the present SiO search,
because the positions of the MSX sources
are more accurately known than those of the IRAS PSC sample and
the AGB nature of the sources was more or less confirmed
by the presence of any NIR counterpart on the 2MASS images near the MSX position.
We conclude that these criteria result in a somewhat flat detection rate with
the flux density.

The average and the standard deviation of the radial velocities of 23 
 MSX sources detected in the $80^{\circ}<l<90^{\circ}$ area 
are $-24.3$ and 25.3 km s$^{-1}$, respectively. 
Because these sources are located 
in the narrow range of $l=85 ^{\circ} \pm 5 ^{\circ}$
where the area is nearly tangential to the direction of the Galactic center, 
the effect of galactic rotation on the radial velocity 
becomes negligible in a range of the distance below 4 kpc [see 
figure 8 of \citet{jia96}]. 
The obtained value of the radial velocity dispersion,
25.3 km s$^{-1}$, is insensitive to the source distances. 
Therefore, it is more accurate than the values obtained before 
for these SiO emitting objects. 
For example,  velocity disperions of $\sim$30 km s$^{-1}$ were obtained
for the southern SiO maser sources in the solar neighbourhood
(\cite{deg01b,nak03a}).

\newpage
\tabcolsep 2pt
\begin{longtable}{crrrrrrrrrl}
\caption{Line parameters for SiO detections.}\label{tab:table1}
\hline\hline
     & \multicolumn{4}{c}{SiO $J=1$--0, $v=1$} & \multicolumn{4}{c}{ SiO $J=1$--0, $v=2$} &  & \\
 \cline{2-5} \cline{6-9}
2MASS name & $V_{\rm lsr}$ & $T_{\rm a}$ & Int. I. & rms 
        & $V_{\rm lsr}$ & $T_{\rm a}$ & Int. I. & rms & $V_{\rm ave}$ & Obs.date \\
     & {\tiny (km s$^{-1}$)} & (K) & {\tiny (K km s$^{-1}$)} & {\tiny (K)} 
 & {\tiny (km s$^{-1}$)} & (K) & {\tiny (K km s$^{-1}$)} & {\tiny (K)}
 & {\tiny (km s$^{-1}$)} & {\tiny (yymmdd.d)} \\
 \hline
\endfirsthead
\hline\hline
     & \multicolumn{4}{c}{SiO $J=1$--0, $v=1$} & \multicolumn{4}{c}{ SiO $J=1$--0, $v=2$} &  & \\
 \cline{2-5} \cline{6-9}
name & $V_{\rm lsr}$ & $T_{\rm a}$ & Int. I. & rms 
     & $V_{\rm lsr}$ & $T_{\rm a}$ & Int. I. & rms & $V_{\rm ave}$ & Obs. date \\
     & {\tiny (km s$^{-1}$)} & (K) & {\tiny (K km s$^{-1}$)} & {\tiny (K)} 
 & {\tiny (km s$^{-1}$)} & (K) & {\tiny (K km s$^{-1}$)} & {\tiny (K)}
 & {\tiny (km s$^{-1}$)} & {\tiny (yymmdd.d)} \\
 \hline
\endhead
\hline
\endfoot
J17575137$-$2431515 & 77.2 & 0.347 & 1.156 & 0.079 & 84.6 & 0.209 & $-$0.134 & 0.091 & 80.9 & 050311.3 \\
J18021839$-$1910397 &   $-$43.5 &  0.412 &   1.379 &  0.062 &   $-$43.3 &  0.416 &   0.947 &  0.079 &   $-$43.4 &  050221.3  \\
J18075037$-$2410216 &    31.4 &  0.274 &   0.732 &  0.053 &    31.4 &  0.358 &   0.864 &  0.066 &    31.4 &  050221.3  \\
J18075221$-$1957160 &   109.9 &  0.839 &   2.655 &  0.094 &   109.9 &  0.845 &   3.035 &  0.101 &   109.9 &  050114.5  \\
J18110459$-$2138129 &   $-$23.7 &  0.582 &   3.004 &  0.083 &   $-$23.5 &  1.141 &   2.480 &  0.120 &   $-$23.6 &  050114.5  \\
J18134615$-$1919197 &    69.8 &  0.309 &   0.404 &  0.055 &    69.5 &  0.578 &   0.798 &  0.077 &    69.6 &  050114.5  \\
J18182915$-$1725379 &   $-$18.4 &  1.005 &   4.420 &  0.071 &   $-$20.1 &  1.124 &   4.085 &  0.096 &   $-$19.3 &  050114.5  \\
J18242192$-$1251548 &    31.3 &  0.459 &   2.846 &  0.133 &    29.7 &  0.701 &   3.161 &  0.158 &    30.5 &  050220.3  \\
J18264227$-$1250409 &    58.1 &  0.441 &   0.540 &  0.053 &    ---  &   ---  &    ---  &  0.068 &    58.1 &  050221.3  \\
J18301610$-$1115376 & 51.4 & 0.412 & 1.655 & 0.065 & 49.9 & 0.292 & 1.041 & 0.057 & 50.7 & 020207.3 \\ 
J18311953$-$0945273 &    43.0 &  0.185 &   0.535 &  0.040 &    --- &  --- &  ---- &  0.055 &    43.0 &  050113.5  \\
J18394573$-$0548423 & 69.7 & 0.097 & 0.381 & 0.029 & 69.5 & 0.145 & 0.452 & 0.029 & 69.6 & 020208.4 \\ 
J18420731$-$0423334 &   104.1 &  0.670 &   4.884 &  0.062 &   104.6 &  0.523 &   2.433 &  0.092 &   104.3 &  050114.5  \\
J18485550$-$0148565 &    10.8 &  2.725 &  11.642 &  0.084 &    10.0 &  2.045 &   8.394 &  0.109 &    10.4 &  050113.5  \\
J18523289+0101103 &    32.5 &  0.175 &   0.386 &  0.051 &    29.2 &  0.392 &   0.258 &  0.074 &    30.9 &  050113.5  \\
J19141961+1110353 &    38.8 &  0.360 &   1.174 &  0.057 &    36.5 &  0.522 &   1.714 &  0.078 &    37.6 &  050114.3  \\
J19574130+3547073 & 19.1 & 0.159 & 0.045 & 0.062 & 20.5 & 0.348 & 0.567 & 0.086 & 19.8 & 040509.1 \\
J20072065+5658529 & $-$1.6 & 0.264 & 0.185 & 0.054 & $-$1.0 & 0.220 & 0.437 & 0.073 & $-$1.3 & 040509.2 \\
J20211407+3537165 & 31.5 & 0.191 & 0.244 & 0.082 & 37.3 & 0.394 & 0.470 & 0.102 & 34.4 & 040516.1 \\
J20224015+5200579 & $-$4.3 & 0.648 & 3.140 & 0.057 & $-$3.2 & 0.380 & 1.469 & 0.077 & $-$3.7 & 040509.2 \\
J20281306+4329251 & $-$29.9 & 0.215 & 0.520 & 0.051 & $-$35.4 & 0.170 & 0.725 & 0.054 & $-$32.7 & 020415.4 \\ 
J20303781+4455348 & $-$8.0 & 0.858 & 2.771 & 0.059 & $-$6.6 & 0.861 & 4.088 & 0.058 & $-$7.3 & 020418.4 \\ 
J20322233+5219419 & $-$70.8 & 2.422 & 5.512 & 0.060 & $-$71.1 & 1.549 & 3.017 & 0.078 & $-$70.9 & 040509.2 \\
J20385860+4505323 & $-$26.4 & 0.614 & 3.617 & 0.067 & $-$23.2 & 0.347 & 1.165 & 0.067 & $-$24.8 & 020418.4 \\ 
J20393950+5012163 & $-$33.0 & 3.086 & 21.550 & 0.091 & $-$38.8 & 3.527 & 18.135 & 0.111 & $-$35.9 & 040526.1 \\
J20415997+4322599 & $-$63.4 & 0.408 & 1.192 & 0.058 & $-$61.4 & 0.342 & 0.788 & 0.050 & $-$62.4 & 020422.3 \\ 
J20432546+4311507 & $-$27.2 & 3.858 & 15.971 & 0.064 & $-$27.9 & 4.282 & 12.188 & 0.071 & $-$27.6 & 020415.4 \\ 
J20441843+5004360 & 19.0 & 0.178 & 0.879 & 0.043 & 23.7 & 0.229 & 0.529 & 0.056 & 21.3 & 040526.1 \\
J20472947+5026369 & $-$29.2 & 1.556 & 4.814 & 0.070 & $-$34.4 & 0.891 & 3.293 & 0.080 & $-$31.8 & 040526.1 \\
J20500334+5020100 & $-$62.2 & 0.654 & 1.863 & 0.103 & $-$61.2 & 0.678 & 1.310 & 0.123 & $-$61.7 & 040526.1 \\
J20504134+3949414 & 1.6 & 1.346 & 5.469 & 0.091 & 2.8 & 1.131 & 4.830 & 0.114 & 2.2 & 020414.4 \\ 
J20533806+4458072 & $-$23.8 & 0.345 & 1.349 & 0.054 & $-$23.9 & 0.594 & 0.963 & 0.053 & $-$23.8 & 020422.3 \\ 
J20571624+4543039 & $-$11.9 & 0.824 & 3.693 & 0.073 & $-$16.0 & 0.621 & 4.248 & 0.075 & $-$13.9 & 020422.3 \\ 
J21012276+4405205 & $-$21.5 & 0.883 & 3.089 & 0.041 & $-$17.0 & 0.486 & 2.433 & 0.041 & $-$19.3 & 020424.3 \\ 
J21020980+4546329 & $-$31.2 & 2.180 & 4.698 & 0.069 & $-$31.3 & 1.159 & 2.794 & 0.064 & $-$31.3 & 020418.4 \\
J21034018+4340193 & 5.2 & 4.939 & 14.962 & 0.437 & 5.0 & 4.180 & 11.354 & 0.328 & 5.1 & 020422.3 \\ 
J21044289+4112450 & $-$7.7 & 0.379 & 0.413 & 0.067 & $-$7.1 & 0.498 & 1.001 & 0.067 & $-$7.4 & 020420.3 \\ 
J21051210+4338462 & $-$33.5 & 0.399 & 1.023 & 0.037 & $-$27.6 & 0.204 & 0.796 & 0.038 & $-$30.5 & 020424.3 \\ 
J21060009+4125316 & $-$52.0 & 0.442 & 1.246 & 0.057 & $-$52.6 & 0.255 & 0.791 & 0.052 & $-$52.3 & 020423.3 \\ 
J21063046+4812550 & $-$57.3 & 0.250 & 0.755 & 0.058 & $-$57.9 & 0.429 & 1.461 & 0.081 & $-$57.6 & 040509.2 \\
J21113676+4151395 & 18.7 & 0.209 & 0.636 & 0.051 & 18.5 & 0.267 & 0.579 & 0.051 & 18.6 & 020420.4 \\ 
J21361612+3231003 &$-$12.7 & 4.791  &20.393 & 0.062   & $-$12.7 & 3.905 & 13.356 & 0.086  & $-$12.7 & 040509.3 \\
OH006.095$-$00.630 & 103.1 & 0.257 & 0.484 & 0.042 & 103.3 & 0.216 & 0.899 & 0.037 & 103.2 & 020207.3 \\
\hline
\end{longtable}
\tabcolsep 4pt
\begin{longtable}{lrrr}
\caption{Negative results for the SiO $J=$1--0 transitions.}\label{tab:table2}
\hline\hline
Name & $v=1$ & $v=2$ & Obs. date \\ 
     & (K) & (K) & (yymmdd.d) \\
\hline
\endfirsthead
\hline\hline
Name & $v=1$ & $v=2$ & Obs. date \\ 
     & (K) & (K) & (yymmdd.d) \\
\hline
\endhead
\hline
\endfoot
J17511085$-$2417088 & 0.109 & 0.084 & 050311.3 \\
J18025777$-$2604009 & 0.058 & 0.044 & 040515.2 \\
J18185116$-$1426388 & 0.091 & 0.061 & 050114.5 \\
J18213815$-$1322056 & 0.080 & 0.055 & 050114.5 \\
J18301278$-$1115463 & 0.086 & 0.078 & 010410.5 \\
J18320111$-$1151333 & 0.095 & 0.067 & 020206.4 \\ 
J18351706$-$0854512 & 0.078 & 0.059 & 050114.5 \\
J18360721$-$0551207 & 0.044 & 0.040 & 020423.3 \\ 
J18361305$-$0718450 & 0.085 & 0.061 & 050114.5 \\
J18362965$-$0544575 & 0.041 & 0.037 & 020414.3 \\ 
J18372528$-$0513452 & 0.067 & 0.061 & 020420.3 \\ 
J18374561$-$0429567 & 0.041 & 0.040 & 020414.3 \\ 
J18374711$-$0146013 & 0.048 & 0.051 & 020425.3 \\
J18375188$-$0549523 & 0.032 & 0.034 & 020208.4 \\ 
J18385495$-$0536293 & 0.030 & 0.034 & 020207.3 \\ 
J18391000$-$0556223 & 0.032 & 0.034 & 020208.4 \\
J18391353$-$0548268 & 0.046 & 0.039 & 020414.3 \\ 
J18391801$-$0548123 & 0.030 & 0.032 & 020207.4 \\ 
J18392571$-$0558011 & 0.035 & 0.037 & 020208.5 \\
J18393640$-$0535362 & 0.027 & 0.029 & 020207.4 \\ 
J18394172$-$0553400 & 0.040 & 0.037 & 020415.3 \\ 
J18482391$-$0252183 & 0.070 & 0.052 & 050113.5 \\
J18504281$-$0021435 & 0.065 & 0.055 & 050113.5 \\
J18561576+0051268 & 0.059 & 0.044 & 050113.5 \\
J19060940+1738007 & 0.052 & 0.043 & 020414.4 \\ 
J19213454+1544256 & 0.038 & 0.034 & 020415.3 \\ 
J19213504+1405168 & 0.062 & 0.060 & 020425.3 \\ 
J19221833+1317598 & 0.057 & 0.061 & 020422.3 \\ 
J19222599+1548008 & 0.047 & 0.039 & 020423.3 \\ 
J19235435+1558391 & 0.130 & 0.143 & 020207.5 \\ 
J19235708+1554214 & 0.030 & 0.034 & 020207.5 \\ 
J19250265+1503486 & 0.039 & 0.043 & 020418.3 \\ 
J19262646+1512434 & 0.059 & 0.063 & 020422.3 \\ 
J19284392+1840220 & 0.038 & 0.034 & 020415.3 \\ 
J19285904+1637014 & 0.073 & 0.052 & 050114.3 \\
J19323648+1905263 & 0.042 & 0.037 & 020415.4 \\ 
J19355481+2014412 & 0.051 & 0.051 & 020208.6 \\ 
J20044596+3611329 & 0.080 & 0.062 & 040509.1 \\
J20051273+3324029 & 0.089 & 0.066 & 040509.1 \\
J20084492+3240237 & 0.090 & 0.069 & 040509.1 \\
J20112621+3542277 & 0.084 & 0.060 & 040509.1 \\
J20113968+3458034 & 0.040 & 0.040 & 020206.6 \\ 
J20125878+3605004 & 0.110 & 0.087 & 040516.1 \\
J20131338+3640303 & 0.096 & 0.078 & 040516.1 \\
J20185949+3528393 & 0.081 & 0.062 & 040509.1 \\
J20203783+4121153 & 0.061 & 0.060 & 020206.6 \\ 
J20230361+3929498 & 0.088 & 0.073 & 040516.1 \\
J20251577+3352539 & 0.082 & 0.063 & 040509.1 \\
J20255437+3732526 & 0.043 & 0.041 & 020208.6 \\ 
J20325512+4232585 & 0.032 & 0.034 & 020207.6 \\
J20340143+4225266 & 0.054 & 0.057 & 020207.6 \\
J20345673+4353406 & 0.062 & 0.067 & 020418.4 \\
J20382595+5141405 & 0.076 & 0.058 & 040509.2 \\
J20385719+4222409 & 0.052 & 0.049 & 020425.3 \\
J20390096+5331336 & 0.075 & 0.056 & 040509.2 \\
J20393745+4252299 & 0.051 & 0.048 & 020415.4 \\
J20400379+4128336 & 0.046 & 0.038 & 020423.3 \\
J20432850+4250018 & 0.053 & 0.055 & 020420.3 \\
J20440953+4056202 & 0.037 & 0.036 & 020424.3 \\
J20451416+4220068 & 0.054 & 0.047 & 020415.4 \\
J20465447+3952182 & 0.054 & 0.053 & 020414.4 \\
J20484510+4059239 & 0.060 & 0.060 & 020207.6 \\
J20513664+4709094 & 0.085 & 0.062 & 040509.1 \\
J20535282+4424015 & 0.061 & 0.068 & 020418.4 \\
J20582798+4738501 & 0.104 & 0.082 & 040516.1 \\
J20585371+4415283 & 0.070 & 0.071 & 020418.4 \\
J21010577+4343012 & 0.051 & 0.049 & 020424.4 \\
J21015501+4517205 & 0.051 & 0.057 & 020422.4 \\
J21020304+4812580 & 0.132 & 0.108 & 040516.1 \\
J21042744+4629511 & 0.143 & 0.117 & 040516.1 \\
J21044605+4634251 & 0.092 & 0.124 & 040526.1 \\
J21072029+4520580 & 0.095 & 0.107 & 020418.4 \\
J21072420+4502462 & 0.045 & 0.039 & 020423.4 \\
J21210028+3813511 & 0.062 & 0.059 & 020425.4 \\ 
J21351887+2731492 & 0.059 & 0.060 & 020425.3 \\  
J21443058+2500260 & 0.051 & 0.048 & 021219.5 \\
J21451786+2227557 & 0.065 & 0.077 & 030521.4 \\ 
J21493147+2241453 & 0.059 & 0.076 & 030521.4 \\ 
OH 012.490$-$00.041  & 0.039 & 0.040 & 020208.3 \\
OH 013.379+00.050    & 0.036 & 0.038 & 020208.3 \\
OH 013.510$-$00.578  & 0.033 & 0.036 & 020208.3 \\
OH 014.431$-$00.033  & 0.035 & 0.035 & 020208.3 \\
OH 018.381+00.162    & 0.035 & 0.036 & 020208.4 \\
OH 028.397+00.080    & 0.053 & 0.054 & 020208.5 \\
OH 031.091$-$00.686  & 0.051 & 0.054 & 020208.5 \\
OH 031.985$-$00.177  & 0.056 & 0.057 & 020208.5 \\
OH 032.731$-$00.327  & 0.055 & 0.059 & 020208.6 \\
\hline
\end{longtable}

\tabcolsep 2pt
\begin{longtable}{lcccccccccc}
\caption{Infrared properties of the observed sources.} 
\hline\hline
\ \ 2MASS name  & \ \ \ $K$ \ \ \ & $J-H$ & $H-K$ & MSX 6C & $\Delta r$ & $F_{\rm C}$ & $C_{\rm AC}$ & $C_{/rm CE}$ & IRAS name & $\Delta r_{\rm I}$ \\
            & (mag) & (mag) & (mag) &          & ($''$)  & (Jy)  &        &          &            & ($''$)  \\
\hline
\endfirsthead
\hline\hline
\ \ 2MASS name  & $K$ & $J-H$ & $H-K$ & MSX 6C & $\Delta r$ & $F_{\rm C}$ & $C_{\rm AC}$ & $C_{\rm CE}$ & IRAS name & $\Delta r_{\rm I}$ \\
            & (mag) & (mag) & (mag) &          & ($''$)  & (Jy)  &        &          &            & ($''$)  \\
\hline
\endhead
\hline
\endfoot
\hline
\multicolumn{11}{l}{$^{\sharp}$ the NIR objects with very faint MSX counterparts.}\\
\multicolumn{10}{l}{$^{\flat}$ 2MASS magnitude with a bad flag.}
\endlastfoot
J17511085$-$2417088     & 8.717 & 3.041 & 2.386  & G004.6251+01.3349  & 1.0 & 11.23 & 0.102 & $-$0.004 &  &  \\ 
J17575137$-$2431515     & 5.531 & 2.655 & 1.490  & G005.1824$-$00.0999  & 1.8 & 10.37 & 0.288 & $-$0.226 &  &  \\ 
J18021839$-$1910397     & 5.644 & 2.144 & 1.408  & G010.3415$+$01.6639  & 1.0 & 13.43 & 0.058 & $-$0.139 &  &  \\ 
J18025777$-$2604009     & 7.545 & 0.515$^{\flat}$ & 2.111$^{\flat}$ & G004.4213$-$01.8638  & 1.2 & 1.51 & $-$0.127 & ---  & & \\
J18075037$-$2410216 & 6.706 & 2.621 & 1.918 & G006.6143$-$01.9021 & 1.1 & 14.54 &  0.117  & $-$0.105 &  &  \\
J18075221$-$1957160 & 7.467 & 5.072 & 2.991 & G010.3061$+$00.1408 & 1.7 & 12.77 &  0.154  & $-$0.018 &  &  \\ 
J18110459$-$2138129 & 4.110 & 1.706 & 0.935 & G009.1956$-$01.3292 & 1.7 & 14.43 & $-$0.00 & $-$0.2037 &  &  \\ 
J18134615$-$1919197 & 6.562 & 3.550 & 2.188 & G011.5293$-$00.7714 & 1.6 & 14.61 &  0.120  & $-$0.079 &  &  \\ 
J18182915$-$1725379 & 5.329 & 2.219 & 1.482 & G013.7286$-$00.8537 & 1.4 & 11.89 &  0.033  & $-$0.114 &  &  \\ 
J18185116$-$1426388 & 8.267 & 4.593 & 3.023 & G016.3981$+$00.4805 & 1.0 & 10.62 &  0.001  & $-$0.201 &  &  \\ 
J18213815$-$1322056 & 5.280 & 3.388 & 1.665 & G017.6654$+$00.3926 & 0.9 & 12.14 &  0.125  & $-$0.287 &  &  \\ 
J18242192$-$1251548 & 3.998 & 1.742 & 1.025 & G018.4213$+$00.0414 & 1.3 & 21.84 &  0.058  & $-$0.310 &  &  \\ 
J18264227$-$1250409 & 2.636 & 2.338 & 1.763 & G018.7056$-$00.4530 & 1.0 & 77.18 &  0.357  & $-$0.269 &  &  \\ 
J18301278$-$1115463$^{\sharp}$ & 6.687 & 2.585 & 1.374  & G020.5053$-$00.4777  & 2.7 & 0.65 & $-$0.097 & --- &  &  \\
J18301610$-$1115376$^{\sharp}$ & 6.233 & 2.435 & 1.438  & G020.5138$-$00.4885  & 4.5 & 2.11 & $-$0.088 & 0.031 &  &  \\
J18311953$-$0945273 & 6.476 & 3.589 & 2.276 & G021.9659$-$00.0234 & 1.1 & 25.43 &  0.158  & $-$0.110 &  &  \\ 
J18320111$-$1151333$^{\sharp}$ & 5.083 & 2.588 & 1.789  & G020.1803$-$01.1454  & 0.6 & 7.82 & 0.053 & $-$0.627 & 18292$-$1153  & 48.2 \\
J18351706$-$0854512 & 5.824 & 3.940 & 2.195 & G023.1652$-$00.5007 & 1.5 & 10.20 &  0.127  & $-$0.112 &  &  \\ 
J18360721$-$0551207$^{\sharp}$ & 7.607 & 2.131 & 1.183  & G025.9766+00.7226  & 0.5 & --- & --- & --- &  &  \\
J18361305$-$0718450 & 4.847 & 3.582 & 1.761 & G024.6935$+$00.0314 & 0.5 & 12.08 &  0.248  & $-$0.397 &  &  \\ 
J18362965$-$0544575$^{\sharp}$ & 7.126 & 2.095 & 1.011  & G026.1137+00.6885  & 0.5 & --- & --- & --- &  &  \\
J18372528$-$0513452$^{\sharp}$ & 7.283 & 3.112 & 1.653  & G026.6816+00.7226  & 0.4 & 0.77 & $-$0.336 & --- &  &  \\
J18374561$-$0429567$^{\sharp}$ & 7.145 & 1.944 & 1.163  & G027.3688+00.9826  & 0.4 & 0.85 & $-$0.042 & --- & 18351$-$0432  & 40.6 \\
J18374711$-$0146013$^{\sharp}$ & 7.024 & 2.127 & 1.214  & G029.8005+02.2293  & 0.6 & --- & --- & --- & 18352$-$0148  & 53.7 \\
J18375188$-$0549523$^{\sharp}$ & 6.077 & 2.358 & 1.197  & G026.1971+00.3480  & 0.5 & 0.79 & $-$0.044 & --- &  &  \\
J18385495$-$0536293$^{\sharp}$ & 6.981 & 3.417 & 1.659  & G026.5155+00.2182  & 0.7 & --- & --- & --- &  &  \\
J18391000$-$0556223$^{\sharp}$ & 6.585 & 2.112 & 0.994  & G026.2489+00.0109  & 0.5 & --- & --- & --- &  &  \\
J18391353$-$0548268$^{\sharp}$ & 7.421 & 2.706 & 1.506  & G026.3735+00.0579  & 0.4 & --- & --- & --- &  &  \\
J18391801$-$0548123$^{\sharp}$ & 6.731 & 2.313 & 1.115  & G026.3858+00.0441  & 0.6 & --- & --- & --- &  &  \\
J18392571$-$0558011$^{\sharp}$ & 5.975 & 1.946 & 0.990  & G026.2547$-$00.0589  & 0.8 & 0.96 & $-$0.176 & --- &  &  \\
J18393640$-$0535362$^{\sharp}$ & 6.920 & 2.206 & 1.078  & G026.6076+00.0722  & 0.5 & --- & --- & --- &  &  \\
J18394172$-$0553400$^{\sharp}$ & 7.183 & 3.392 & 1.692  & G026.3458$-$00.0821 & 17.3 & 0.00 & --- & --- &  &  \\
J18394573$-$0548423$^{\sharp}$ & 5.834 & 1.984 & 1.117  & G026.4310$-$00.0619  & 0.8 & 1.67 & $-$0.150 & --- & 18370$-$0551  & 54.6 \\
J18420731$-$0423334 & 6.552 & 6.208 & 3.183 & G027.9618$+$00.0652 & 1.3 & 13.93 &  0.295  & $-$0.277 &  &  \\ 
J18482391$-$0252183 & 4.186 & 2.246 & 0.989 & G030.0296$-$00.6339 & 1.8 & 18.11 &  0.158  & $-$0.183 &  &  \\ 
J18485550$-$0148565 & 3.792 & 1.461 & 0.969 & G031.0297$-$00.2690 & 0.9 & 16.13 &  0.064  & $-$0.284 &  &  \\ 
J18504281$-$0021435 & 4.372 & 1.702 & 0.775 & G032.5274$-$00.0036 & 1.8 & 11.92 &  0.106  & $-$0.220 &  &  \\ 
J18523289$+$0101103 & 6.009 & 3.128 & 1.800 & G033.9660$+$00.2178 & 0.7 & 10.78 &  0.102  & $-$0.198 &  &  \\ 
J18561576$+$0051268 & 8.830 & 4.102 & 2.237 & G034.2456$-$00.6821 & 2.1 & 11.76 &  1.844  &    --- & & \\ 
J19060940+1738007$^{\sharp}$   & 7.520 & 2.774 & 2.093  & G050.3014+04.8022 & 0.4 & 6.23 & 0.068 & --- & 19039+1733  & 13.1 \\
J19141961$+$1110353 & 7.821 & 3.081 & 2.750 & G045.4712$+$00.0762 & 0.2 & 12.66 &  0.148  &  0.104 &  &  \\ 
J19213454+1544256$^{\sharp}$   & 7.259 & 2.741 & 1.398  & G050.3295+00.6522  & 0.4 & --- & --- & --- &  &  \\
J19213504+1405168$^{\sharp}$   & 7.696 & 2.792 & 1.356  & G048.8739$-$00.1273  & 0.2 & --- & --- & --- & 19192+1359  & 14.5 \\
J19221833+1317598$^{\sharp}$   & 7.845 & 2.225 & 1.137  & G048.2598$-$00.6534  & 0.2 & --- & --- & --- &  &  \\
J19222599+1548008$^{\sharp}$   & 7.206 & 2.346 & 1.123  & G050.4795+00.4980  & 0.4 & --- & --- & --- &  &  \\
J19235435+1558391$^{\sharp}$   & 6.938 & 2.161 & 0.984  & G050.8034+00.2689  & 0.4 & --- & --- & --- &  &  \\
J19235708+1554214$^{\sharp}$   & 6.606 & 2.681 & 1.281  & G050.7453+00.2262  & 0.3 & --- & --- & --- &  &  \\
J19250265+1503486$^{\sharp}$   & 7.179 & 2.644 & 1.328  & G050.1280$-$00.4046  & 0.3 & --- & --- & --- &  &  \\
J19262646+1512434$^{\sharp}$   & 7.609 & 2.231 & 1.099  & G050.4190$-$00.6305  & 0.5 & --- & --- & --- &  &  \\
J19284392+1840220$^{\sharp}$   & 7.167 & 2.452 & 1.276  & G053.7217+00.5382  & 0.5 & --- & --- & --- &  &  \\
J19285904$+$1637014 & 7.067 & 2.932 & 1.477 & G051.9458$-$00.4976 & 2.4 & 10.95  &  1.546  &    --- & & \\ 
J19323648+1905263$^{\sharp}$   & 7.043 & 2.411 & 1.272  & G054.5285$-$00.0649  & 0.4 & --- & --- & --- &  &  \\
J19355481+2014412$^{\sharp}$   & 6.355 & 2.312 & 1.430  & G055.9154$-$00.1858  & 0.5 & 1.34 & 0.064 & --- &  &  \\
J19574130+3547073       & 5.470 & 1.085 & 0.762  & G071.7632+03.4869 & 1.4 & 5.72 & 0.047 & $-$0.217 & 19558+3538 & 10.0 \\
J20044596+3611329       & 3.805 & 1.052 & 0.623  & G072.8639+02.4845 & 1.7 & 9.66 & 0.030 & $-$0.091 & 20028+3602 & 14.3 \\
J20051273+3324029       &13.928 & 0.747$^{\flat}$ & 2.974$^{\flat}$ & G070.5537+00.9140 & 3.3 & 13.11 & 0.120 & $-$0.104 & 20032+3315 & 1.9 \\
J20072065+5658529       & 2.097 & 0.987 & 0.547 & G091.0070+12.9818 & 1.3 & 40.33 & 0.142 & $-$0.311 & 20062+5650 & 3.6 \\
J20084492+3240237       & 11.740 & 0.929$^{\flat}$ & 3.432 & G070.3404$-$00.1023 & 1.5 & 13.51 & 0.140 & $-$0.176 & 20067+3231 & 4.0 \\
J20112621+3542277       & 2.602 & 1.214 & 0.647  & G073.1873+01.0867 & 2.0 & 16.59 & 0.022 & $-$0.167 & 20095+3533 & 2.7 \\
J20113968+3458034$^{\sharp}$   & 6.580 & 1.890 & 1.022  & G072.5933+00.6429  & 0.4 & 1.08 & $-$0.309 & --- & 20098+3449  & 58.9 \\
J20125878+3605004       & 6.047 & 2.274 & 1.266  & G073.6728+01.0322  & 3.0 & 8.27 & 0.093 & 0.026 & 20110+3555  & 4.8 \\
J20131338+3640303       & 3.669 & 1.338 & 0.691  & G074.1937+01.3176  & 3.3 & 18.31 & 0.131 & $-$0.162 & 20113+3631  & 1.7 \\
J20185949+3528393       & 7.802 & 4.401 & 3.182  & G073.8496$-$00.3173 & 1.5 & 34.22 & 0.086 & $-$0.191 & 20171+3519 & 24.7 \\
J20203783+4121153$^{\sharp}$   & 5.141 & 2.713 & 1.456  & G078.8804+02.7394  & 0.5 & 2.14 & 0.134 & --- &  &  \\
J20211407+3537165       & 1.623 & 1.256 & 0.404  & G074.2263$-$00.6124  & 1.8 & 38.50 & 0.004 & $-$0.239 & 20193+3527  & 16.2 \\
J20224015+5200579       & 2.985 & 1.019 & 1.030  & G087.9334+08.4578 & 1.4 & 25.19 & $-$0.010 & $-$0.327 & 20212+5151 & 4.6 \\
J20230361+3929498       & 4.679 & 1.179 & 1.121  & G077.6160+01.3026  & 4.6 & 41.76 & 0.040 & $-$0.013 & 20212+3920  & 9.5 \\
J20251577+3352539       & 4.548 & 1.668 & 0.867  & G073.2741$-$02.2867 & 0.9 & 11.96 & 0.111 & $-$0.120 & 20233+3343 & 2.7 \\
J20255437+3732526$^{\sharp}$   & 6.745 & 2.264 & 1.051  & G076.3424$-$00.2750  & 0.7 & 0.69 & $-$0.249 & --- & 20239+3722  & 43.6 \\
J20281306+4329251 & 3.503 & 1.454 & 0.884  & G081.4357+02.8203  & 4.8 & 39.48 & $-$0.048 & $-$0.225 & 20264+4319  & 3.5 \\
J20303781+4455348 & 2.661 & 1.256 & 0.784  & G082.8561+03.3092  & 1.4 & 12.20 & 0.042 & $-$0.383 &  &  \\
J20322233+5219419 & 3.305 & 1.226 & 0.804  & G089.0500+07.4149 & 0.1 & 9.98 & $-$0.091 & $-$0.186 &  &  \\
J20325512+4232585 & 5.828 & 3.621 & 2.782  & G081.1810+01.5718  & 4.7 & 64.45 & 0.015 & $-$0.268 & 20311+4222  & 7.4 \\
J20340143+4225266 & 1.013 & 2.073 & 0.856  & G081.2011+01.3332  & 2.4 & 149.20 & $-$0.046 & $-$0.150 & 20322+4215  & 6.1 \\
J20345673+4353406 & 7.521 & 3.828 & 2.744  & G082.4817+02.0772  & 0.2 & 18.80 & 0.051 & $-$0.356 &  &  \\
J20382595+5141405 & 7.874 & 2.248 & 1.900  & G089.0919+06.2882 & 0.6 & 53.88 & 0.079 & $-$0.167 & 20369+5131 & 1.7 \\
J20385719+4222409 & 8.213 & 2.835 & 2.134  & G081.7131+00.5792  & 2.4 & 16.30 & $-$0.246 & 0.650 &  &  \\
J20385860+4505323 & 3.250 & 0.945 & 0.911  & G083.8714+02.2257  & 0.4 & 13.90 & 0.017 & $-$0.411 &  &  \\
J20390096+5331336 & 2.220 & 0.962 & 0.480  & G090.6207+07.3163 & 0.7 & 20.44 & 0.027 & $-$0.285 & 20376+5320 & 3.7 \\
J20393745+4252299 & 7.739 & 3.158 & 2.423  & G082.1824+00.7839  & 4.9 & 20.03 & 0.010 & $-$0.283 &  &  \\
J20393950+5012163 & 3.709 & 0.997 & 0.613  & G088.0140+05.2359  & 1.6 & 52.02 & 0.019 & $-$0.152 & 20381+5001  & 1.4 \\
J20400379+4128336 &11.786 & 2.859 & 2.324  & G081.1225$-$00.1343  & 2.0 & 5.70 & $-$0.064 & 0.670 &  &  \\
J20415997+4322599 & 5.639 & 2.004 & 1.124  & G082.8495+00.7526  & 5.3 & 5.20 & $-$0.093 & $-$0.206 &  &  \\
J20432546+4311507 & 1.491 & 1.266 & 0.487  & G082.8629+00.4337  & 1.9 & 72.00 & 0.034 & $-$0.370 &  &  \\
J20432850+4250018 &12.120 & 2.609$^{\flat}$ & 3.116$^{\flat}$ & G082.5828+00.2014  & 0.2 & 6.55 & $-$0.353 & 0.308 &  &  \\
J20440953+4056202 & 6.441$^{\flat}$ & 1.505 &$-$2.104$^{\flat}$ & G081.1745$-$01.0720  & 1.2 & 6.30 & $-$0.053 & $-$0.380 &  &  \\
J20441843+5004360 & 4.354 & 1.360 & 0.767 & G088.3694+04.5685  & 2.5 & 6.35 & 0.013 & $-$0.209 & 20427+4953  & 1.6 \\
J20451416+4220068 & 1.731 & 1.112 & 0.667 & G082.3932$-$00.3620  & 1.2 & 15.24 & 0.057 & $-$0.351 &  &  \\
J20465447+3952182 & 2.431 & 1.369 & 0.839 & G080.6650$-$02.1460  & 1.6 & 9.05 & 0.111 & $-$0.261 &  &  \\
J20472947+5026369 & 4.155 & 1.460 & 0.955 & G088.9757+04.3973  & 1.2 & 13.67 & $-$0.001 & $-$0.196 & 20459+5015  & 4.2 \\
J20484510+4059239 & 2.044 & 0.965 & 0.506 & G081.7560$-$01.7163  & 2.4 & 13.65 & 0.022 & $-$0.285 &  &  \\
J20500334+5020100 & 7.826 & 2.495 & 1.734 & G089.1487+04.0113  & 2.4 & 4.68 & 0.094 & 0.043 & 20484+5008  & 2.6 \\
J20504134+3949414 & 2.297 & 1.234 & 0.812 & G081.0888$-$02.7370  & 1.6 & 18.27 & 0.011 & $-$0.265 &  &  \\
J20513664+4709094 & 5.419 & 1.703 & 1.168 & G086.8479+01.7937 & 1.9 & 17.53 & 0.032 & $-$0.092 & 20499+4657 & 3.5 \\
J20533806+4458072 &13.468 & 0.764$^{\flat}$ & 3.748$^{\flat}$ & G085.3935+00.1268  & 3.0 & 17.40 & $-$0.201 & 0.198 &  &  \\
J20535282+4424015 & 3.304 & 1.868 & 0.979  & G084.9864$-$00.2715  & 1.2 & 44.22 & $-$0.075 & $-$0.271 &  &  \\
J20571624+4543039 & 4.838 & 1.790 & 1.176  & G086.3782+00.1237  & 2.1 & 19.63 & $-$0.060 & $-$0.065 &  &  \\
J20582798+4738501 & 2.799 & 1.188 & 0.720  & G087.9748+01.2270  & 0.7 & 15.78 & $-$0.001 & $-$0.231 & 20567+4727  & 4.0 \\
J20585371+4415283 & 6.228 & 1.014 & 0.807  & G085.4597$-$01.0466  & 1.1 & 7.99 & $-$0.172 & 0.303 &  &  \\
J21010577+4343012 & 6.470 & 2.417 & 1.660  & G085.3119$-$01.7005  & 2.2 & 6.65 & 0.056 & $-$0.404 &  &  \\
J21012276+4405205 & 4.838 & 1.396 & 0.850  & G085.6254$-$01.4932  & 2.0 & 10.47 & $-$0.017 & $-$0.086 &  &  \\
J21015501+4517205 & 2.924 & 1.106 & 0.638  & G086.5890$-$00.7718  & 0.8 & 16.72 & $-$0.003 & $-$0.217 &  &  \\
J21020304+4812580 & 5.202 & 3.000 & 2.408  & G088.7983+01.1488  & 3.2 & 57.41 & $-$0.046 & $-$0.302 & 21003+4801  & 5.6 \\
J21020980+4546329 & 3.166 & 1.347 & 1.191  & G086.9824$-$00.4822  & 1.0 & 21.09 & 0.035 & $-$0.242 & \multicolumn{2}{l}{(=V407 Cyg)}  \\
J21034018+4340193 & 3.377 & 1.475 & 1.039  & G085.5869$-$02.0784  & 3.2 & 21.82 & $-$0.045 & $-$0.274 &  &  \\
J21042744+4629511 & 9.899 & 3.027 & 2.793  & G087.7861$-$00.2986  & 3.9 & 11.60 & 0.055 & $-$0.223 & 21026+4617  & 18.3 \\
J21044289+4112450 & 2.874 & 1.163 & 0.746  & G083.8783$-$03.8607  & 4.1 & 7.48 & 0.015 & $-$0.441 &  &  \\
J21044605+4634251 & 7.244 & 1.701 & 1.079  & G087.8784$-$00.2874  & 3.5 & 2.47 & 0.133 & $-$0.160 & 21030+4622  & 7.6 \\
J21051210+4338462 & 3.707 & 1.080 & 0.772  & G085.7522$-$02.3024  & 4.5 & 9.13 & 0.013 & $-$0.362 &  &  \\
J21060009+4125316 & 4.381 & 1.283 & 0.778  & G084.1989$-$03.8987  & 5.3 & 13.54 & $-$0.020 & $-$0.207 &  &  \\
J21063046+4812550 & 5.530 & 1.328 & 0.822  & G089.2934+00.5954 & 2.2 & 5.89 & 0.038 & $-$0.107 & 21048+4800 & 4.4 \\
J21072029+4520580 & 6.420 & 2.875 & 2.418  & G087.2711$-$01.4405  & 4.5 & 53.32 & $-$0.016 & $-$0.301 &  &  \\
J21072420+4502462 & 7.675 & 2.567 & 2.147  & G087.0547$-$01.6535  & 4.3 & 11.49 & 0.049 & $-$0.397 &  &  \\
J21113676+4151395 & 2.778 & 1.264 & 0.839  & G085.2311$-$04.3761  & 6.5 & 9.26 & 0.070 & $-$0.424 &  &  \\
J21210028+3813511 & 4.952 & 0.871 & 0.476  & (=V1903 Cyg)        &      &      &       &  &  & \\
J21351887+2731492 & 6.715 & 0.873 & 0.496  & (=BR Peg)           &      &      &       &  &  & \\
J21361612+3231003 & 1.975 & 0.963 & 0.630  & G082.0070$-$14.4349 & 1.9  &33.85 & 0.001 & 0.341 &  &  \\
J21443058+2500260 & 4.114 & 0.764 & 0.393  & (=RR Peg)           &      &      &       &  &  & \\
J21451786+2227557 & 6.565 & 0.878 & 0.450  & (=CO Peg)           &      &      &       &  &  & \\
J21493147+2241453 & 6.738 & 0.853 & 0.438  & (=CX Peg)           &      &      &       &  &  & \\
\end{longtable}

\tabcolsep 3pt
\begin{longtable}{lrrrrcrrcc}
\caption{List of observed OH 1612 MHz sources [from \citet{sev01}]}\label{tab:table5}
\hline\hline
OH name &\multicolumn{2}{c}{ RA \  (J2000.0) \  Dec } 
& $V_{\rm c}$ & $V_{\rm out}$ & MSX6C & Sep & $F_{\rm A}$  & Nearest IRAS & Type \\
\cline{2-3}
 &  h \ m \ s \ & $\circ$ \ \ $'$ \ \ $''$ \ & {\tiny (km s$^{-1}$)} & {\tiny (km s$^{-1}$)} &
 & ($''$)   & (Jy) & Obj. and sepr.($'$) & \\
\endfirsthead
\hline
\endhead
\hline
\endfoot
\hline
\endlastfoot
\hline
OH 006.095$-$00.630 & 18 01 51.8 &$-$24 00 08.5 & 102.2  & 13.6 & G006.0952$-$00.6294 & 2.0 & 1.59 & 17588$-$2358 (2.04) &   star \\
OH 012.490$-$00.041 & 18 13 00.5 &$-$18 07 44.9 &  63.5 &  13.6 & ---                 & --- &  --- & 18099$-$1808 (1.58) &  ?SF \\
OH 013.379+00.050   & 18 14 27.5 &$-$17 18 17.1 &  52.2 &  13.6 & G013.3797+00.0499   & 1.7 & 6.45 & 18114$-$1718 (1.23) &  star \\
OH 013.510$-$00.578 & 18 17 02.0 &$-$17 29 22.8 &  36.3  &  2.2 & ---                 & --- & ---  & 18140$-$1726 (3.86) &  star \\
OH 014.431$-$00.033 & 18 16 51.4 &$-$16 25 11.0 &  77.2 &   4.6 & G014.4311$-$00.0328   & 1.6 & 0.30 & 18141$-$1626 (1.98) &  star \\
OH 018.381+00.162   & 18 23 50.9 &$-$12 50 41.5 &  20.5  &  9.1 & ---                 & --- & ---  & 18211$-$1251 (1.57) &  ?star \\
OH 028.397+00.080   & 18 42 51.9 &$-$03 59 54.6 &  74.9  & ---  & G028.3937+00.0757   &20.8 & 1.79 & 18402$-$0403 (0.81) &  ?SF \\
OH 031.091$-$00.686 & 18 50 31.4 &$-$01 57 05.2 &  98.6 &  19.3 & ---                 & --- & ---  & 18476$-$0158 (5.23) &  ?SF \\
OH 031.985$-$00.177 & 18 50 20.5 &$-$00 55 24.6 & 111.1 &  11.4 & ---                 & --- & ---  & 18478$-$0058 (0.92) &  star \\
OH 032.731$-$00.327 & 18 52 14.1 &$-$00 19 39.6 &  60.2 &  14.8 & ---                 & --- & ---  & 18494$-$0022 (3.55) &  ?star \\
\hline
\end{longtable}

\begin{longtable}{lccrrrcc}
  \caption{Result of NIR observations of OH maser sources.}\label{tab:table8}
  \hline\hline
Name    &  \multicolumn{2}{c}{RA \   (J2000.0) \    Dec}   &     $J$  &    $H$  &  $K$   &  SiO & MSX6C \\
\cline{2-3}
        &  h \ m \ s  \ & \ \ $\circ$ \ \ $' $ \ \ $''$ \ & (mag) & (mag) & (mag) & /H$_2$O & \\
\endfirsthead
\hline
OH 006.095$-$00.630 & 18 01 51.90  & $-$24 00 08.4   &  ---   & ---  &  13.760  & y  & y \\
OH 012.490$-$00.041 & \multicolumn{5}{c}{Not observed. The object not found on the 2MASS images.} & n  &  n   \\
OH 013.379+00.050 & 18 14 27.56 & $-$17 18 16.7&  13.870 &  9.997  &  $<$8.574  & n  &     y   \\
OH 013.510$-$00.578 & 18 17 02.06 &$-$17 29 22.8 & 12.434  & $<$9.836 &  $<$9.222 & n &   n   \\
OH 014.431$-$00.033 & 18 16 51.48 &$-$16 25 10.9 & 16.627 & 13.764  & 12.221  & n  & y   \\
OH 018.381+00.162 & \multicolumn{2}{c}{Object not found } &  ---   & ---    & ---   & n & n  \\  
OH 028.397+00.080 & \multicolumn{2}{c}{Object not found }  &  ---   & ---    & ---  &  y &  ?   \\
OH 031.091$-$00.686 & \multicolumn{2}{c}{Object not found }  &  ---   & ---    & ---  &   n &  n   \\
OH 031.985$-$00.177 & 18 50 20.63 &$-$00 55 23.8 &  13.018 & 10.403  & $<$9.355  &  n & n \\
OH 032.731$-$00.327 & \multicolumn{2}{c}{Object not found}   &  ---   & ---    & ---   &  n   &  n  \\
\hline
 \multicolumn{8}{l}
{The inequality, $<$, indicates that the photon count is saturated.}\\
\end{longtable}
\tabcolsep 1pt
\begin{longtable}{lrrrrrrrrc}

  \caption{Results of MIR observations for NIR and OH sources.}\label{tab:table9}
  \hline\hline
Name      & $F_{8.7}$ & $F_{9.8}$ & $F_{12.5}$ & $F_{24.5}$ & $T_{\rm dust}$ & $C_{12}$ & 
$A_K^{\ddagger}$ & $D_{L}$ & $D_{\rm K}$ \\
          &   (Jy)    &   (Jy)    &   (Jy)   &   (Jy)   &   (K)     &   & (mag) & 
          (kpc) &(kpc)  \\
\endfirsthead
\hline    
J18301610$-$1115376     & 2.30    & 2.60    & 2.78    & & 404 & $-0.19$ &0.86&  4.5 & 3.6\\   
J18394573$-$0548423     & 1.48    & 1.92    & 2.17    & & 344 & $-0.10$ &0.96&  3.8 & 4.1\\   
OH006.095$-$00.630       & 1.78    & 0.18    & 2.60    & & 345 & $-0.10$ &1.09&  8.2 & 9.2$^{\dagger}$ \\
OH012.490$-$00.041       & $<0.11$ & $<0.22$ & $<0.16$ & $<1.1$ & --- &  ---    &--- & $>10$ & 11.3$^{\dagger}$\\
OH013.379+00.050         & 5.21    &  3.17   & 8.41    & & 320 & $-0.05$ &1.13&  4.4 & 4.9\\
OH013.510$-$00.578       & 0.18    & $<0.13$ & $<0.24$ & & 371 & $-0.14$ &1.74&  9.6 & 11.3$^{\dagger}$\\
OH014.431$-$00.033       & 0.34    & 0.17    & 1.17    & & 211 & 0.31    &3.47&  8.1 & 10.4$^{\dagger}$\\
OH018.381+00.162         & $<0.11$ & $<0.17$ & 0.61    & 3.5 & --- &  ---    &--- & $>10$ & 13.2$^{\dagger}$\\  
OH028.397+00.080         & $<0.07$ & $<0.14$ & $<0.17$ & $<1.5$ & --- &  ---    &--- & $\sim 5$ & 11.1$^{\dagger}$ \\
OH031.091$-$00.686       & $<0.07$ & $<0.27$ & $<0.14$ & $<1.4$ & --- &  ---    &--- & $>10$ & 8.0$^{\dagger}$ \\
OH031.985$-$00.177       & 2.83    & 3.13    & 3.12    & & 442 & $-0.23$ &1.63&  6.0 & 6.8\\
OH032.731$-$00.327       & $<0.10$ & $<0.23$ & $<0.22$ & $<1.1$ & --- &  ---    &--- & $>10$ & 9.8$^{\dagger}$ \\
\hline 
 \multicolumn{10}{l}
{The inequalities in the 2--5th columns indicate that the source was not detected. }\\
 \multicolumn{10}{l}
{$^{\dagger}$ The kinematic distance at the far side of the tangential point.}\\
 \multicolumn{10}{l}
 {$^{\ddagger}$ Interstellar extinction used for the luminosity-distance computation.}\\  
\end{longtable}
%
\clearpage
\begin{figure*}[t]
\begin{center}
\leavevmode
\input epsf
\epsfverbosetrue
\epsfxsize 10mm 
\end{center}
\caption{ 
a. SiO $J=1$--0 $v=1$ and 2 spectra of the detected sources. 
The source name and observing date are shown on the upper left of each panel.  
}
\label{fig1a}
\end{figure*} 
\setcounter{figure}{0}
\begin{figure*}[t]
\begin{center}
\leavevmode
\input epsf
\epsfverbosetrue
\epsfxsize 10mm 
\caption{b. same as figure 1a.}
\label{fig1b}
\end{center}
\end{figure*} 
\setcounter{figure}{0}
\begin{figure*}[t]
\begin{center}
\leavevmode
\input epsf
\epsfverbosetrue
\epsfxsize 10mm 
\caption{c. same as figure 1a, except the final panel 
of OH 028.397+0.080 for the 22 GHz H$_2$O spectrum.}
\label{fig1c}
\end{center}
\end{figure*} 

\clearpage
\setcounter{figure}{1}
\begin{figure*}
  \begin{center}
\leavevmode
\input epsf
\epsfverbosetrue
\epsfxsize 10mm 
  \caption{$JHK$ color-composite images of OH maser sources. 
  The size is $128 \times 128 ''$; north is up and east is left.
  The identified sources are indicated by the arrow.}\label{fig2}
\end{center}
\end{figure*}
%
\begin{figure*}
  \begin{center}
\leavevmode
\input epsf
\epsfverbosetrue
\epsfxsize 10mm 
  \caption{Two-color (left) and magnitude--color (right) diagrams
 ($J-H$ and $K$ plots against $H-K$) for the NIR sources (magnitudes taken from the 2MASS database). 
  The large filled circle indicates the SiO detection 
  (J18301610$-$1115376 and J18394573$-$0548423) and the open circles indicate  nondetections. 
 The square indicates an M5III star at 8 kpc away without extinction.
 It is expected to move along the broken line in this diagram with 
 interstellar and circumstellar reddening. The area enclosed by thick broken lines
 indicates the approximate region of SiO maser stars found in the galactic bulge (\cite{deg01a}).
 }
\label{fig3}
\end{center}

\end{figure*}
\clearpage
\begin{figure*}
  \begin{center}
\leavevmode
\input epsf
\epsfverbosetrue
\epsfxsize 10mm 
  \caption{
 Observed spectra of the NIR (upper panel) and OH 1612 MHz (lower panel) sources.
 The observed points are at wavelengths, 1.25($J$), 1.65($H$), 
 2.20($K$), 8.78, 9.73, and 12.41 $\mu$m.
 The upper panel shows the spectra of the two NIR sources with SiO detections
 and the lower panel the spectra for OH 1612 MHz sources.
  }
\label{fig4*}
\end{center}

\end{figure*}
\begin{figure*}
  \begin{center}
\leavevmode
\input epsf
\epsfverbosetrue
\epsfxsize 10mm 
  \caption{
 Histograms of $F_{12}$ and $C_{\rm CE}$ for the MSX sources.
 The shaded areas indicate SiO detection and white non-detection.
 The line graphs indicate the detection rate 
  and the scales are shown on the right.
  }
\label{fig5}
\end{center}
\end{figure*}


\begin{thebibliography}{}
\bibitem[Becker et al.~(1994)]{bec94}
Becker R. H., White R. L., Helfand D. J., \& Zoonematkermani S. 1994, ApJS, 91, 347
\bibitem[Beichman et al.~(1988)]{bei88}
Beichman C. A., Neugebauer G., Habing H. J., Clegg P. E., \&
Chester T. J. 1988, IRAS Catalogs and Atlases, I. Explanatory
Supplement, NASA RP-1190 (Washington D.C.; US Government Printing Office) 
\bibitem[Bowers, de Jong~(1983)]{bow83}
Bowers, P. F., \& de Jong, T. 1983, AJ, 88, 655 
\bibitem[Carey et al. (2000)]{car00} 
Carey, S. J., Feldman, P. A., Redman, R. O., Egan, M. P., MacLeod, J. M., \& 
Price, S. D. 2000, ApJ, 543, L157
\bibitem[Caswell~(1999)]{cas99} Caswell, J. L., 1999, MNRAS, 308, 683
\bibitem[Cohen et al.~(1999)]{coh99}
Cohen, M., Walker, R. G., Carter, B., Hammersley, P., Kidger, M., \& Noguchi, K.
1999, AJ, 117, 1864

\bibitem[Deguchi et al.~(2000)]{deg00a}
Deguchi, S., Fujii, T.,  Izumiura, H., Kameya, O., 
  Nakada, Y., Nakashima, J., Ootsubo, T., \& Ukita, N. 2000, ApJS, 128, 571 
  
\bibitem[Deguchi et al.~(2001a)]{deg01a}
Deguchi, S., Fujii, T., Matsumoto, S.  Nakashima, J., \& Wood, P. 2001a, PASJ, 53, 293 
    
\bibitem[Deguchi et al.~(2002)]{deg02}
Deguchi, S., Fujii, T., Nakashima, J., \& Wood, P. R., 2002, PASJ, 54, 719 

\bibitem[Deguchi et al.~(2004a)]{deg04a}
Deguchi, S., et al.
2004a, PASJ, 56, 261

\bibitem[Deguchi et al.~(2004b)]{deg04b}
Deguchi, S. et al.
2004b, PASJ, 56, 765

\bibitem[Deguchi et al.~(1998)]{deg98}
Deguchi S., Matsumoto S., \& Wood P. R. 1998, PASJ, 50, 597

\bibitem[Deguchi et al.~(2001b)]{deg01b} 
Deguchi, S. Nakashima, J., \& Balasubramanyam, R. 2001b, PASJ, 53, 305

\bibitem[Egan et al.~(1999)]{ega99}
Egan M. P., et al.
1999, Air Force Research Laboratory Technical Report
No. AFRL-VS-TR-1999-1522 
(available at IPAC, $\langle$http://irsa.ipac.caltech.edu/applications/MSX/$\rangle$
 with the MSX catalog)

\bibitem[Engels~(2002)]{eng02}
Engels, D. 2002, A\&A, 388, 252

\bibitem[Fix, Mutel~(1984)]{fix84}
Fix, J. D., \& Mutel, R. L. 1984, AJ, 89, 406 

\bibitem[Gaume, Mutel~(1987)]{gau87} Gaume, R. A., \& Mutel, R. L. 1987, ApJS, 65, 193

\bibitem[Glass et al.~(2001)]{gla01}
Glass, I. S., Matsumoto, S., Carter, B. S.,  \& Sekiguchi, K. 2001, MNRAS 321, 77 ;Erratum: 336, 1390

\bibitem[Herman et al.~(1985)]{her85}
Herman, J., Baud, B., Habing, H. J., \& Winnberg, A. 1985,
A\&A, 143, 122

\bibitem[Izumiura et al.~(1999)]{izu99}
Izumiura H.,
Deguchi S., Fujii T., Kameya O., Matsumoto S., Nakada Y., 
Ootsubo T., \& Ukita N. 1999,  ApJS, 125, 257

\bibitem[Jewell et al.~(1991)]{jew91}
Jewell, P. R., Snyder, L. E., Walmsley, C. M., Wilson, T. L., \& Gensheimer, P. D.
1991, A\&A, 242, 211

\bibitem[Jiang~(2002)]{jia02}
Jiang, B. W. 2002, ApJ, 566, L37

\bibitem[Jiang et al.~(1997)]{jia97}
Jiang, B. W., Deguchi, S., Hu, J.-Y., Yamashita, T., Nishihara, E., 
  Matsumoto, S., \& Nakada, Y. 1997, AJ, 113, 1315 

\bibitem[Jiang et al.~(1999)]{jia99}
Jiang, B. W., Deguchi, S., \& Ramesh, B. 1999, PASJ, 51, 95

\bibitem[Jiang et al.~(1996)]{jia96}
Jiang, B. W., Deguchi, S.,
  Yamamura, I., Nakada, Y., Cho, S. H.,  \& Yamagata, T. 1996, ApJS, 106, 463

\bibitem[Johansson et al.~(1977)]{joh77}
Johansson, L. E. B., Andersson, C., Goss, W. M., \& Winnberg, A. 1977, A\&AS, 28, 199

\bibitem[Kataza et al.~(2000)]{kat00}   Kataza, H. ,  Okamoto, Y.,  Takubo, S.
Onaka, T.,  Sako, S.,  Nakamura, K.,  Miyata, T.,  \& Yamashita, T. 2000, SPIE, 4008, 1144
\bibitem[Lumsden et al.~(2002)]{lum02}
Lumsden, S. L.,  Hoare, M. G.,  Oudmaijer, R. D., \&  Richards, D. 2002, 
MNRAS, 336, 621
 
\bibitem[Minter et al.~(2001)]{min01}
Minter, A. H., Lockman, F. J., Langston, G. I., \& Lockman, J. A. 2001, ApJ, 555, 868

\bibitem[Miyata et al.~(2000)]{miy00}
Miyata, T., Kataza, H., Okamoto, Y., Onaka, T., \& Yamashita, T. 2000, ApJ, 531, 917

\bibitem[Munari et al.~(1990)]{mun90}
Munari, U., Margoni, R., \& Stagni, R. 1990, MNRAS, 242, 653

\bibitem[Nagashima et al.~(1999)]{nag99}
Nagashima, C., et al.
1999, Proc. Intern. Conf. ``Star Formation 1999'',
 ed. T. Nakamoto,  (Nagano; Nobeyama Radio Observatory), p394 

\bibitem[Nagayama et al.~(2003)]{nag03}
Nagayama, T., et al.
 2003, SPIE, 4841, 459

\bibitem[Nakashima, Deguchi~(2003a)]{nak03a}
Nakashima, J., \& Deguchi, S., 2003a, \pasj, 55, 203

\bibitem[Nakashima, Deguchi~(2003b)]{nak03b}
Nakashima, J., \& Deguchi, S., 2003b, \pasj, 55, 229

\bibitem[Patel et al.~(1992)]{pat92}
Patel, N. A., Joseph, A., \& Ganesan, R. 1992, JApA, 13, 241

\bibitem[Seaquist et al.~(1995)]{sea95}
Seaquist, E. R., Ivison, R. J. \& Hall, P. J. 1995, MNRAS, 276, 867
\bibitem[Sevenster~(2002)]{sev02}
Sevenster, M. 2002, AJ, 123, 2788
\bibitem[Sevenster et al.~(1997)]{sev97b}
Sevenster M. N., Chapman J. M., Habing H. J., Killeen, N. E. B., \& Lindqvist M.
1997, A\&AS, 122, 79
\bibitem[Sevenster et al.~(2001)]{sev01}
Sevenster M. N., van Langevelde, H. J. Moody, R. A. Chapman J. M., Habing H. J., \& Killeen, N. E. B.
2001, A\&A, 366, 481

\bibitem[Skrutskie et al.~(2000)]{skr00}  
Skrutskie, M. F.,  Stiening, R., Cutri, R., Beichman, C., Capps, R.,  
Carpenter, J., Chester, J., Elias, J. et al.\ 2000 (The 2MASS Team)\\ 
http://www.ipac.caltech.edu/2mass/overview/2massteam.html

\bibitem[Sun, Zhang (1998)]{sun98}
Sun, J., \& Zhang, H-Y. 1998, Chin. Astron. Astrophys., 22, 442
\bibitem[Takaba et al.~(2001)]{tak01}
Takaba, H., Iwata, T., Miyaji, T., \& Deguchi, S. 2001, PASJ, 53, 517
\bibitem[Tatarnikova et al. (2003)]{tat03}	
Tatarnikova, A. A., Marrese, P. M., Munari, U.,  Tomov, T., Whitelock, P. A., \& Yudin, B. F. 2003, MNRAS, 344, 1233
\bibitem[te Lintel Hekkert et al.~(1991)]{tel91}
te Lintel Hekkert, P., Caswell, J. L., Habing, H. J., Haynes, R. F., \&
Norris, R. P. 1991, A\&AS, 90, 327

\bibitem[van Hoof et al.~(1997)]{van97}
van Hoof, P. A. M., Oudmaijer, R. D., \& Waters, L. B. F. M 1997, 
MNRAS, 289, 371
\bibitem[Winberg et al.~(1975)]{win75}
Winnberg, A., Nguyen-Quang-Rieu,  Johansson, L. E. B., \& Goss, W. M. 1975, A\&A, 38, 145
\bibitem[Zijlstra et al.~(1990)]{zil90}
Zijlstra, A. A., Pottasch, S. R., Engels, D., Roelfsema, P. R., te Lintel Hekkert, P. \& Umana, G.
1990, MNRAS, 246, 217
\end{thebibliography}
\end{document}